\documentclass[11pt,english]{article}
\usepackage[sc]{mathpazo}

\usepackage{geometry}
\usepackage{color}
\usepackage{array}
\usepackage{bbding}
\usepackage{multirow}
\usepackage[fleqn]{amsmath}
\usepackage{amssymb}
\usepackage{amsthm}
\usepackage{graphicx}

\usepackage{lscape}

\usepackage{natbib}

\usepackage{comment}
\usepackage{bm}
\usepackage[title]{appendix}
\usepackage{longtable}
\usepackage{mathtools}
\usepackage{chngcntr}
\usepackage{authblk}
\usepackage{algorithm}
\usepackage{algorithmicx}
\usepackage{algpseudocode}
\usepackage{amsmath}
\usepackage{url}
\usepackage{diagbox}
\usepackage{longtable}

\usepackage[colorlinks,
            linkcolor=blue,
            anchorcolor=blue,
            citecolor=blue
            ]{hyperref}

\makeatletter
\allowdisplaybreaks

\@ifundefined{showcaptionsetup}{}{%
 \PassOptionsToPackage{caption=false}{subfig}}
\usepackage{subfig}
\makeatother

\usepackage{babel}
\usepackage{enumitem}

\let\textcolor\relax 

\begin{document}\sloppy

\title{Vehicle Dispatching and Routing of On-Demand Intercity Ride-Pooling Services: A Multi-Agent Hierarchical Reinforcement Learning Approach}
\author{Jinhua Si\textsuperscript{a}\hspace{2em} Fang He\textsuperscript{a}\footnote{Corresponding author. E-mail address: \textcolor{blue}{fanghe@tsinghua.edu.cn}.}\hspace{2em} Xi Lin\textsuperscript{b}\hspace{2em} Xindi Tang\textsuperscript{c}}
\affil{\small\emph{\textsuperscript{a}Department of Industrial Engineering, Tsinghua University, Beijing 100084, P.R. China}\normalsize}
\affil{\small\emph{\textsuperscript{b}Department of Civil and Environmental Engineering, University of Michigan, Aann Arbor 48109, United States}\normalsize}
\affil{\small\emph{\textsuperscript{c}School of Management Science and Engineering, Central University of Finance and Economics, Beijing 100081, P.R. China}\normalsize}
\date{\today}
\maketitle

\begin{abstract}
\noindent The integrated development of city clusters has given rise to an increasing demand for intercity travel. Intercity ride-pooling service exhibits considerable potential in upgrading traditional intercity bus services by implementing demand-responsive enhancements. Nevertheless, its online operations suffer the inherent complexities due to the coupling of vehicle resource allocation among cities and pooled-ride vehicle routing. To tackle these challenges, this study proposes a two-level framework designed to facilitate online fleet management. Specifically, a novel multi-agent feudal reinforcement learning model is proposed at the upper level of the framework to cooperatively assign idle vehicles to different intercity lines, while the lower level updates the routes of vehicles using an adaptive large neighborhood search heuristic. Numerical studies based on the realistic dataset of Xiamen and its surrounding cities in China show that the proposed framework effectively mitigates the supply and demand imbalances, and achieves significant improvement in both the average daily system profit and order fulfillment ratio.


\end{abstract}

\section{Introduction} \label{sec_introduction}
\noindent Conventional intercity bus services are facing increasing operational pressures. The number of intercity bus users in many countries has declined significantly in recent years (\citealp{GANJI2021345}). The scale of the intercity bus industry in China, where more than 70$\%$ passenger trips are delivered by road transport, has been shrinking over the past decade. A large number of bus stations in China shut down due to continued losses, and the nationwide number of passenger vehicles has decreased by 30.06$\%$ compared to 2015 (\citealp{tr2021}). The intercity bus industry in the United States also began shrinking after 2015. Greyhound, which operates the largest intercity bus service in North America, reduced its operations by 16$\%$ between 2016 and 2020 (\citealp{schwieterman2021brink}). The possible reasons \textcolor{}{are twofold}. From the \emph{supply} perspective, the steady increase in car ownership compresses the demand for short-haul bus lines, while the infrastructure construction of high-speed railways and airlines tends to be improved in more regions, delivering fatal blows to many longer-haul bus lines. From the \emph{demand} perspective, people's travel behaviors have been reshaped in the mobile Internet era, \textcolor{}{placing increased emphasis on convenience, comfort, and privacy when traveling. The attractiveness of traditional intercity buses is diminishing due to heightened competition from alternative modes of intercity transportation}. \par

\begin{figure}
  \centering
  \includegraphics[width=0.8\linewidth]{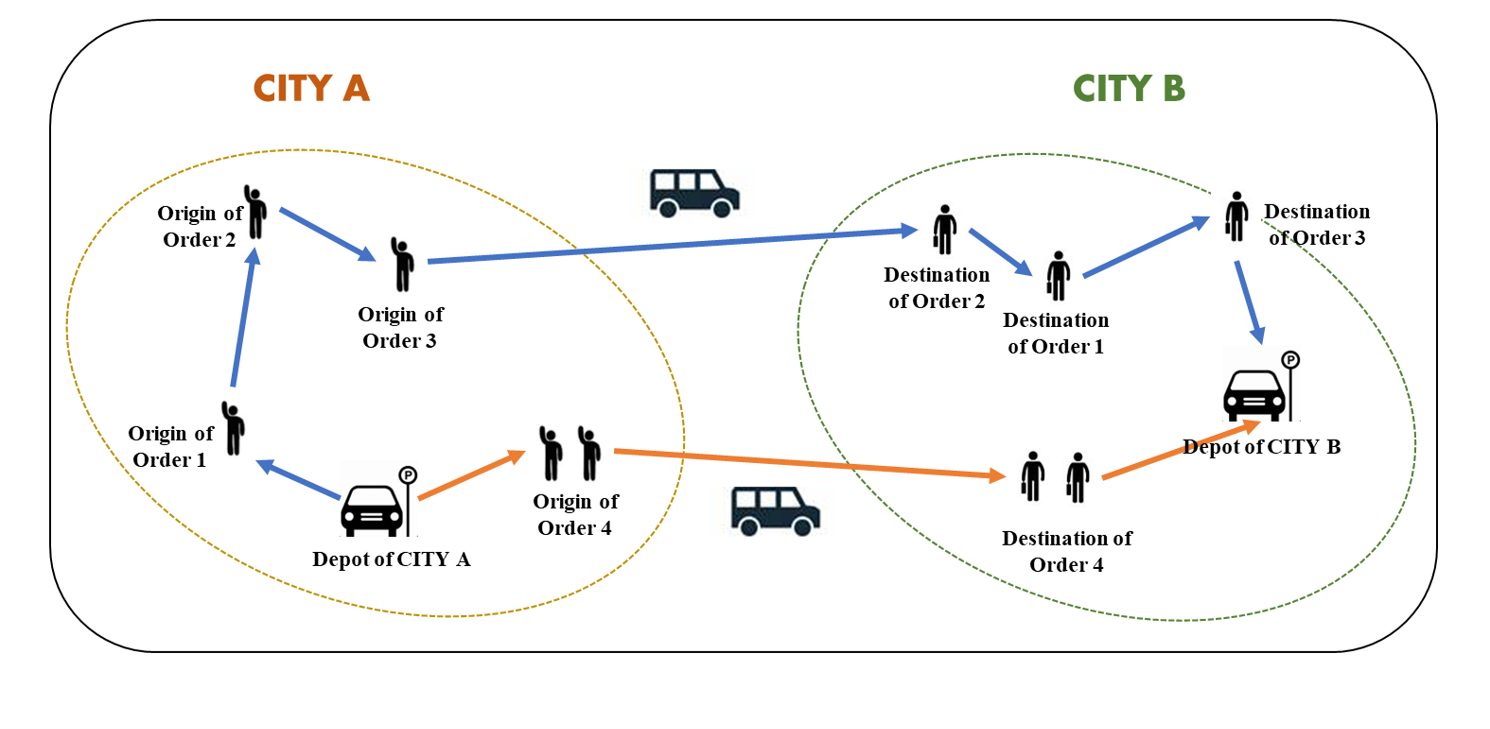}
  \caption{Intercity passenger transport in a pooled-ride manner.}
  \label{fig:exp_poolride}
\end{figure}

\noindent With the development of mobile Internet and the proliferation of smartphones, the intercity ride-pooling service within city clusters is becoming a promising direction for the digital transition and demand-responsive upgrading of the conventional intercity bus service. As an emerging travel mode, it utilizes vehicles \textcolor{}{with smaller capacities to provide door-to-door service for each passenger}. Similar to urban ride-hailing platforms, a computing center makes centralized decisions in real time based on the information of orders and available vehicles. Each order is expected to specify the number of passengers, the locations for pick-up and drop-off in different cities, the time windows, and other necessary information. The computing center assigns orders to vehicles precisely and plans the route for each vehicle to pick up passengers in the departure city and deliver them to their destinations in a pooled-ride manner, as shown in Figure \ref{fig:exp_poolride}.

\noindent The advantages of intercity ride-pooling services in city clusters are mainly manifested in three aspects. 
The \emph{first} advantage is replacing the traditional fixed-schedule service with a \textcolor{}{dynamic} ride-hailing service. The demand-responsive transit system, which has demonstrated its remarkable effectiveness in significantly enhancing the travel experience within urban public transport over the last decade (\citealp{wang2019ridesourcing}; \citealp{liu2021mobility}; \citealp{vansteenwegen_survey_2022}), holds great promise for delivering a more convenient and diversified travel experience in intercity transport. The demand-responsive services can also enhance system efficiency by matching supply and demand more intelligently and efficiently. 
The \emph{second} advantage lies in the ability to plan fully flexible routes for each vehicle instead of fixed routes, offering door-to-door services to passengers. With the gradual expansion of urban areas and increased traffic congestion, the generalized cost for passengers concentrating at bus stations in the city has increased dramatically, which makes door-to-door services increasingly appealing and advantageous. 
\textcolor{}{The \emph{third} advantage is that it fully considers the allocation of transportation resources at the city cluster level}. The highly developed city cluster has become an increasingly significant geographic unit under the rapid urbanization process, where the strong socioeconomic connections between the cores and the peripheries brew up strong intercity travel demand (\citealp{fang2017urban}). Road passenger transport is still competitive within city clusters, especially for those clusters with relatively low railway coverage. Rather than being subordinated to a fixed line between two cities, vehicles can be dispatched to any city within the cluster according to demand fluctuation, which can resolve the spatial imbalance between supply and demand. \par

\noindent While the aforementioned advantages highlight a promising market for intercity ride-pooling services within city clusters, there are still several issues that hinder the development of this innovative service. The core is how to build a highly effective and intelligent fleet operations strategy for the sophisticated dynamic system and unique scenarios of intercity travel. From the perspective of macroscopic spatiotemporal allocation of vehicle resources, platforms need to implement dynamic dispatching of fleets among multiple cities to reduce the gap between available seats and potential demand between each city pair in future periods. Extensive explorations have been conducted in the operational strategy optimization in related fields, such as urban ride-hailing and ride-sourcing platforms(\citealp{xu2018large}; \citealp{Tang2020}; \citealp{guo2021robust}; \citealp{kullman2022dynamic}). From the perspective of microscopic order-matching and vehicle routing, the platform needs to solve the multi-vehicle pooled-ride routing problem to maximize the profit of each route in an online manner, which is a variant of the Pickup and Delivery Problem with Time Windows (PDPTW) or Dial-a-Ride Problem (DARP) (\citealp{Cordeau2006}; \citealp{Ropke2006}). 
The fleet operations strategy is based on the integrated optimization of macroscopic fleet dispatching and microscopic pooled-ride vehicle routing, which has been barely touched on in existing literature under intercity scenarios. The coupling structure makes it difficult to devise an efficient fleet management strategy using model-based approaches. 
Additionally, the differences between intracity and intercity travel lead to different optimization frameworks for ride-pooling services. The intercity demand-responsive transport system faces greater challenges in coping with supply and demand dynamics due to longer travel times.  Additionally, relocation of idle vehicles is generally unreasonable in intercity ride-pooling services due to high travel costs. The intercity travel demand within a city cluster can be clustered into several mutually independent intercity lines, and the platform has to allocate vehicle resources to lines across broader temporal and spatial dimensions.

\noindent \textcolor{}{This study proposes an online two-level framework to dispatch vehicles among several cities effectively and update intercity pooled-ride vehicle routes dynamically}. The \emph{dispatching level} (upper level) assigns idle vehicles to intercity lines based on global demand and supply information. The dispatching decisions for each city are derived synergistically using a novel multi-agent hierarchical reinforcement learning method, named multi-agent feudal networks (MFuN), to enhance agent cooperation and far-sighted assignment. The \emph{routing level} (lower level) uses the adaptive large neighborhood search heuristic (ALNS) to solve the order-vehicle matching and pooled-ride routing problems at each time interval based on the decisions of the dispatching level. The two-level framework can provide online operational decisions while avoiding myopic actions under the guidance of deep reinforcement learning. The system aims to improve both the system profit and order fulfillment ratio, enabling service for more passengers within their reserved time windows. The main contributions of this study can be summarized as follows.
\begin{itemize}
    \item This study focuses on the online fleet operations of on-demand intercity ride-pooling services, while the applications of demand-responsive services in the realm of intercity transport have received scarce research attention thus far. The coupling of vehicle resource allocation among cities and multi-vehicle pooled-ride routing problems brings challenges to proposing an integrated model for dispatching and routing. We also take into account the distinctive attributes of intercity travel when developing strategies, including long travel times and high intercity costs. 
    
    \item We propose an integrated online operational framework for intercity ride-pooling services within a city cluster. The framework comprises two levels, with the upper level dedicated to addressing vehicle resource allocation among cities in the long term, and the lower level focusing on dynamic routing. At each dispatching horizon, the upper level of the framework optimizes the distribution of vehicle resources within a city cluster by dispatching vehicles to intercity lines to mitigate the spatiotemporal imbalances between supply and demand. At each matching interval with a shorter time scale, the lower level solves dynamic DARP for each intercity line respectively to enable efficient online management of the fleet at the scale of city clusters. 
    
    \item We take advantage of a novel multi-agent hierarchical reinforcement learning method, named MFuN, in the upper level of the framework, to develop a non-myopic strategy for the stochastic dynamic vehicle resource allocation problem. MFuN shows favorable performance in promoting agent cooperation and capturing transition patterns in system states over extended horizons, and ensures that the actions of each agent contribute to maximizing the long-term profits of the entire system.
    
    \item We acquire realistic operational data of intercity ride-pooling services to evaluate the effectiveness of our framework. The numerical experiments conducted reveal that the proposed framework can develop proactive operational strategies to alleviate the spatiotemporal imbalance between supply and demand, leading to remarkable performance enhancements on both system profit and order fulfillment ratio.
\end{itemize}

\noindent The remainder of this paper proceeds as follows. Section \ref{sec_literature} reviews recent literature addressing DARP and fleet dispatching problems. Section \ref{sec_description} specifically portrays the modeling of the problem. Section \ref{sec_methodology} introduces the proposed two-level operational framework. Section \ref{sec_experiment} conducts numerical experiments on two toy networks and one realistic network to demonstrate the optimization effectiveness under various supply and demand scenarios, and provide managerial insights for the platform operations. Finally, we summarize the study and provide an outlook on future research in section \ref{sec_conclusion}.

\section{Literature Review} \label{sec_literature}

\noindent \textcolor{}{As an emerging service mode motivated by increasingly customized travel preferences and growing intercity population movements, intercity ride-pooling services remain largely unexplored in the literature.} Nevertheless, researchers have made several progress in solving related problems, such as the pooled-ride routing problem and urban demand-responsive fleet dispatching problem.

\noindent Multi-vehicle pooled-ride routing problem is a variant of the Pickup and Delivery Problem with Time Windows (PDPTW) or Dial-a-Ride Problem (DARP) (\citealp{Ho2018}). This problem can be formulated as a mixed-integer linear program (MILP) problem, and existing solution methods can be categorized into heuristic methods (\citealp{Ropke2006}; \citealp{naccache2018multi}; \citealp{goeke2019granular}; \citealp{guo2022vehicle}) and exact methods(\citealp{Cordeau2006}; \citealp{gschwind2015effective}; \citealp{braekers2016multi}; \citealp{luo2019two}). However, for large-scale problems, exact algorithms such as branch-and-cut algorithms and branch-and-price algorithms are usually unable to provide optimal or even feasible solutions within an acceptable time, so heuristic algorithms such as neighborhood search algorithms \textcolor{}{are more widely applied in solving practical problems}. Adaptive large neighborhood search (ALNS), first proposed by \cite {Ropke2006}, is one of the most effective heuristic algorithms to solve large-scale multi-vehicle dial-a-ride problems with pick-up and delivery time windows. ALNS utilizes multiple removal and insertion operators to enhance the movement in solution space in a simulated annealing metaheuristic manner. Subsequent studies have proposed many detailed techniques to improve the performance of the algorithm or to solve different variants of DARP (\citealp{Demir2012}; \citealp{Ghilas2016}; \citealp{Gschwind2019}; \citealp{SunP2020}). 

\noindent \textcolor{}{The aforementioned routing algorithms for DARP are widely used in the fleet operations of urban demand-responsive services such as flexible bus systems and ride-pooling services. 
The rest of this section reviews studies in vehicle routing and fleet dispatching of these systems. We also consider several papers that focus on ride-hailing services with no shared rides, as their approaches can be effectively extended to ride-pooling services. 
}

\noindent Flexible bus systems in practice can be classified into two categories. 
The first category is semi-flexible that each bus travels along a standard route and can deviate from the predetermined route to serve emerging demand, while the second category is fully flexible that routes and timetables are determined based on demand (\citealp{vansteenwegen_survey_2022}). Some studies model the semi-flexible bus scheduling and routing as static problems, among which the majority combine candidate stops into routes based on demand spatial distribution (\citealp{czioska2019real}; \citealp{sun2020optimizing}), while others collaboratively optimize the selection of bus stops and suggestions for pick-up locations (\citealp{melis2022static}). Matching of new orders and dynamic adjustment of routes are considered to develop dynamic models in some literature ( \citealp{fielbaum2021demand}; \citealp{tafreshian2021proactive}; \citealp{melis2022real}). In the literature of fully flexible bus systems, the online vehicle routing problems are usually modeled as deterministic dynamic DARP and solved by the methods mentioned above(\citealp{agatz2012optimization}; \citealp{braekers2016multi}; \citealp{huang2020two}; \citealp{guo2022vehicle}; \citealp{wu2022time}). Certain literature uses stochastic optimization to capture the uncertainty of demand distribution and detour time (\citealp{lee2021zonal}). 

\noindent \textcolor{}{The fleet operation problem for urban ride-hailing services constitutes one of the most extensively investigated topics of demand-responsive transport research. 
Although the majority of these studies predominantly concentrate on ride matching and vehicle dispatching with no shared ride, many studies have explored the operation problem of urban ride-pooling systems. Some studies proposed solution frameworks for the ride-pooling systems with requests in advance(\citealp{maNearondemandMobilityBenefits2022a}), while others focused on the on-demand request matching and real-time vehicle routing.
\citet{alonso-moraPredictiveRoutingAutonomous2017} and \citet{simonettoRealtimeCityscaleRidesharing2019} recast the real-time operation problem into a series of batch processes for request assignment and vehicle routing. The batch matching optimization for non-pooling services can be formulated as a bipartite matching problem (\citealp{tong2016online}; \citealp{qin2021multi}). However, because of the time window constraints for successive pick-up and delivery, the batch-matching process is more complicated when multiple requests can be served by the same vehicle (\citealp{ouyang2021performance}).}
Some literature proposes a dynamic optimization model to determine the optimal matching interval and radius (\citealp{gyh2020}; \citealp{yang2020optimizing}; \citealp{qin2021optimizing}). Reinforcement learning methods for dynamic matching also have gained immense attention in recent years to prevent short-sighted decisions (\citealp{xu2018large}; \citealp{tang2019deep}; \citealp{qin2020ride}).

\noindent Based on the aforementioned batch-matching framework, the spatiotemporal imbalance between supply and demand can be further mitigated by proactive vehicle dispatching. By leveraging extensive operational data, the platform can develop a sophisticated supply and demand prediction model, offering guidance for real-time vehicle dispatching (\citealp{ke2017short}; \citealp{guo2020residual}). Methods for fleet dispatching in urban ride-hailing services can be broadly classified into two categories. The \emph{first} category models the problem as a Markov decision problem and constructs various optimization models(\citealp{xu2018large}). 
\textcolor{}{Various methods are developed to provide pooled-ride vehicle routing and rebalancing results under this stochastic environment, including shareability graph-based algorithm (\citealp{alonso-moraPredictiveRoutingAutonomous2017}, \citealp{tuncelIntegratedRidematchingVehiclerebalancing2023}), approximate dynamic programming approach(\citealp{yuIntegratedDecompositionApproximate2020}),  linear assignment algorithm (\citealp{simonettoRealtimeCityscaleRidesharing2019} ), and data-driven metaheuristics(\citealp{bongiovanni2022machine}).}
\citet{Lei2020} propose a two-layer framework, where the upper layer focuses on dynamic pricing and relocation, while the lower layer determines the passengers' choice behavior based on market equilibrium. \citet{guo2021robust} employ a rolling-horizon robust optimization approach based on model predictive control (MPC) for vehicle location optimization. 

\noindent The \emph{second} category adopts model-free approaches to tackle large-scale fleet operations for ride-hailing services. Many studies use single-agent reinforcement learning to provide centralized decisions. \citet{Tang2020} designs a two-layer single-agent reinforcement learning framework to solve the problem of online dispatching, charging arrangement, and order matching for urban electric vehicles. \citet{liu2022deep} propose a single-agent deep reinforcement learning approach for the vehicle dispatching problem, in which a single-agent queue is designed for dispatching multiple vehicles based on a global pruned action space. More studies utilize multi-agent reinforcement learning (MARL) framework to derive dispatching and scheduling strategies based on the well-trained value functions (\citealp{xu2018large}; \citealp{jin2019coride}; \citealp{guo2020deep}; \citealp{jiao2021real}; \citealp{kullman2022dynamic}). In these studies, an agent is usually defined as a vehicle or vehicles within a grid, with eligible actions including repositioning to a certain area and fulfilling certain orders. The innovations of these studies are mainly focused on applying frontier reinforcement learning techniques to enhance the model's effectiveness, such as the use of attention mechanisms and graph neural networks. Compared to single-agent reinforcement learning methods, multi-agent frameworks attain more efficient performance in large-scale problems, but their policies are decentralized and may sacrifice system-level optimality (\citealp{mao2020dispatch}). 

\noindent To summarize, although demand-responsive fleet management has garnered sustained attention in recent years, the distinctive characteristics of intercity ride-pooling services still pose challenges to the direct application of existing frameworks. Specifically, for the routing problem of each ride-pooling vehicle, the existing literature has proposed comprehensive methods for effective computation. However, applying routing algorithms at the city cluster level can result in a vast search space, while introducing intercity lines can simplify the multi-vehicle pooled-ride routing problem at the city cluster level into separate vehicle routing within each line. For fleet dispatching, the aforementioned dispatching strategies for urban flexible bus systems and ride-hailing services are not directly applicable to fleet operations within a network comprising multiple cities. Intercity transport is based on a topology of city networks with relatively long duration for intercity trips, and relocation of idle vehicles is generally not cost-effective. The most formidable challenge resides in the coupling between long-term vehicle resource allocation and dynamic vehicle routing, which has received scant attention in existing research. Both intercity dispatching and dynamic pooling are difficult to solve using model-based approaches, while realistic fleet operations necessitate an efficient online framework to integrate the optimization of these two levels.

\section{Modeling} \label{sec_description}

\subsection{Problem settings}\label{subsec_setting}
\noindent In our study, we consider fleet operations in the context of intercity ride-pooling services within a city cluster. The cluster $\mathcal{U}=\{1,...,n\}$ contains several interconnected cities, and the platform manages a fleet to fulfill intercity travel demands within the cluster in a pooled-ride manner. Multiple orders served by the same vehicle in one ride must originate from the same city and travel to the same city. It is assumed in this study that orders with different origin cities or different destination cities can not be simultaneously served by the same vehicle because visiting other cities en route will significantly increase the detour time for passengers on board. Under this assumption, the ride-pooling services within the city cluster can be split into services for a set of intercity lines $\mathcal{L}\subset \mathcal{U}\times\mathcal{U}$, in which each line $\ell \in \mathcal{L}$ starts from city ${o^\ell \in \mathcal{{U}}}$ and ends at city ${d^\ell \in \mathcal{{U}}}$ without visiting any intermediate city. Notice that ride-pooling service is not available for all pairs of cities within the cluster. The platform exclusively offers intercity lines between cities with sufficient demand for intercity travel, whereas the demand level between other city pairs can be so minimal that ride-pooling services cease to benefit from economies of scale. Furthermore, the lines in this study are different from the fixed lines of conventional intercity bus services. Conventional intercity buses travel along fixed routes on a fixed schedule, while vehicles in this study travel along fully flexible routes on each line. Instead of adhering to specific lines, each vehicle can be dynamically assigned to any line originating from its current city for upcoming rides.

\noindent From the \emph{supply} perspective, the fleet dispatching is based on the directed graph $G(\mathcal{U},\mathcal{L})$. We assume all vehicles in the fleet $\mathcal{K}$ are identical in capacity and speed, but they originate from different cities and commence their operations at different times to work continuously for no more than a preset time limit $\bar{W}$. The platform assigns orders to vehicles so that vehicles can pick up and deliver passengers in the planned sequence to complete a trip. Since intercity trips are relatively costly, relocation between cities is discouraged, and hence the platform only assigns idle vehicles to lines originating from the cities they are located. Furthermore, for the sake of long-distance transportation safety, drivers have to take a certain period of rest at the depot after finishing an intercity ride.\par

\noindent \textcolor{}{From the \emph{demand} perspective, this study primarily focuses on providing services to `on-demand' intercity trip orders, wherein passengers require service within a relatively short period after reservation (e.g., within 2 hours). In urban ride-hailing services, the on-demand order represents that the passenger expects prompt service immediately upon placing the order. However, when travelers need intercity trips, they tend to formulate travel arrangements in advance and anticipate receiving services within a predetermined time window. 
We can usually observe significant reservations at the early stage of the days-long booking period (\citealp{tsai2020self}), partly owing to the fact that traditional intercity transit system faces challenges in handling demand fluctuations caused by `on-demand' orders with short booking lead-time. 
In our study, necessary information of each order includes the number of passengers, origin, destination, hard time windows, and the selected line. Passengers may also have requirements for the latest arrival time at their destinations. Passengers are assumed to wait at the designated origins within the reserved time windows for pick-up, so they will not leave the system until they get picked up or the reserved time window ends. In other words, we do not model the behavior of passengers canceling orders due to excessive waiting time in this study, because passengers can express their travel preferences by personalizing the time windows. Actually, our framework can be easily extended to address the problems with soft time window constraints or order cancellations. In our study, a feasible passenger-vehicle matching necessitates a compatible vehicle arriving at the origin and destination location before the respective latest time. The matching process is executed at short intervals for each line respectively. Once the order is assigned to a certain vehicle during one matching process, the passenger will receive the information of the matched vehicle, so neither the driver nor the passenger can cancel the matched order. If no feasible match can be made before the latest pick-up time, the platform will forfeit this order.}\par

\noindent \subsection{Modeling framework} \label{framework}
In this subsection, we briefly introduce the overall decision-making framework in fleet dispatching and vehicle routing of the intercity ride-pooling services. We discretize the daily time domain into $T$ dispatching horizons of equal duration and the set of all horizons is denoted as $\mathcal{T}$. Orders of each line emerge following a specific stochastic process throughout the day, which are added to the matching pool of the respective line in real-time. The fleet operations within a dispatching horizon $t\in \mathcal{T}=\{ 0, 1, ..., T-1\}$ contain two levels: \emph{first}, the platform assigns idle vehicles to different lines or reserves them at their current cities; \emph{second}, the platform plans a route for each vehicle to fulfill orders in the matching pool for each line respectively. \textcolor{}{These two levels of fleet operations are executed at different temporal resolutions. The first-level idle fleet dispatching decisions are made once in a dispatching horizon, while the matching and routing are updated more frequently to improve order response rate and avoid order loss. Therefore, the platform splits each dispatching horizon into several matching intervals with equal lengths and updates the planned routes of all vehicles en route at each matching interval.}

\begin{figure}[!ht]
  \centering
  \includegraphics[width=1.0\linewidth]{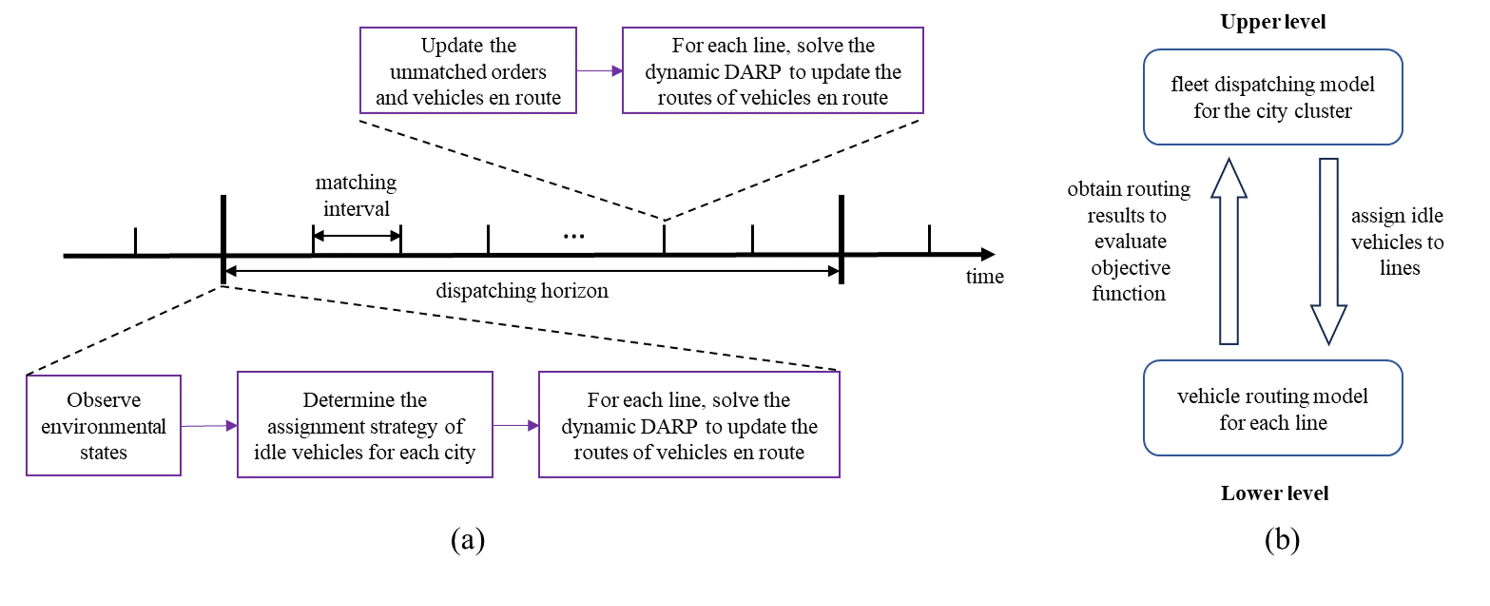}
  \caption{(a) Illustration of the dispatching horizons and matching intervals. (b) Interaction between two levels.}
  \label{fig:framework}
\end{figure}

\noindent \textcolor{}{The relationship between dispatching horizons and matching intervals is further illustrated in Figure \ref{fig:framework} (a).}
At the beginning of a dispatching horizon, the platform develops dispatching decisions for the current horizon based on the global supply and demand information. A relatively short horizon may lead to no idle vehicles for assignment, while an excessively long horizon could compromise the effectiveness of the dispatching. Based on realistic operational data and numerical tests, we set the horizon duration as 20 minutes in this study. A horizon is composed of a set of batch-matching processes. The first batch matching is executed instantly after the assignment of idle vehicles, which can provide initial routes for vacant vehicles to fulfill orders, and the other batch matching \textcolor{}{processes in a horizon are} executed at the beginning of each matching interval. The matching interval can be within the range of 0.5 to 2 minutes in practice. \par

\noindent Based on this operational framework, the intercity fleet operations problem can be regarded as a two-level embedded fleet dispatching and vehicle routing problem. The vehicle routing problem at the lower level is based on the decisions \textcolor{}{of assigning} vehicles to lines derived from the upper level, and the objective function value of the fleet dispatching problem at the upper level is obtained by the routing results of the lower-level problem, as illustrated in Figure \ref{fig:framework} (b). \par

\subsection{The fleet dispatching model}\label{dispatching}
\noindent The fleet dispatching problem at the upper level of the modeling framework can be described as a stochastic dynamic resource allocation problem (SDRAP) (\citealp{godfrey2002adaptive}). Let $\hat {\mathcal{K}}^t_{u}$ denote the set of vehicles that enter the system to start working at the beginning of horizon $t\in \mathcal{T}$ at city $u\in \mathcal{U}$, $\tilde {\mathcal{K}}^t_u$ denote the set of vehicles that exit the system due to reaching limited work duration at the beginning of horizon $t$ at city $u$, $\check {\mathcal{K}}^t_{u}$ denote the set of arrived vehicles at horizon $t$ at city $u$, $\bar {\mathcal{K}}^{t}_{u}$ denote the set of reserved vehicles at horizon $t$ at city $u$. \textcolor{}{We assume that all drivers have to take a rest of at least $\tau_r$ horizons after an intercity trip. 
Then all available vehicle resources for assignment at horizon $t$ at city $u$ is ${\mathcal{K}}^{t}_u=\hat {\mathcal{K}}^t_{u} \cup \check {\mathcal{K}}^{t-\tau_r-1}_{u} \cup \bar {\mathcal{K}}^{t-1}_{u}\setminus \tilde{\mathcal{K}}^t_{u}$.}
At horizon $t$, we determine the spatial allocation of idle vehicles $\mathcal N^t = (\mathcal N_{uv}^t)_{(u,v)\in \mathcal L}$, where $\mathcal N_{uv}^t$ is the set of vehicles dispatched from city $u$ to city $v$ at horizon $t$. 
Then the dynamics of all available vehicles in each city over horizons can be represented by the flow conservation constraints (\ref{vehicle_dynamic_0}),
(\ref{vehicle_dynamic_1}), (\ref{vehicle_dynamic_t}) and (\ref{vehicle_dynamic_en_route}), where $\tau_{uv}$ in constraint (\ref{vehicle_dynamic_en_route}) denotes the expected travel time between city $u$ and city $v$. Additionally, it is anticipated that vehicles will return to their initial city once their daily working time reaches the designated limit $\bar W$, as shown in constraint (\ref{vehicle_work_limit}).
\begin{align}
    &\hat {\mathcal{K}}^t_{u} = (\bigcup_{v:(u,v)\in\mathcal{L}}\mathcal N_{uv}^t) \cup \bar {\mathcal{K}}^t_u & t=0,\forall u\in\mathcal{U}\label{vehicle_dynamic_0}\\
    &\textcolor{}{
    \hat {\mathcal{K}}^t_{u} \cup \bar {\mathcal{K}}^{t-1}_u \setminus \tilde{\mathcal{K}}^t_{u}= (\bigcup_{v:(u,v)\in\mathcal{L}} \mathcal N_{uv}^t) \cup \bar {\mathcal{K}}^t_u}
    &\textcolor{}{
    \forall t\in \{1,\cdots ,\tau_r\}, u\in\mathcal{U}}\label{vehicle_dynamic_1}\\ 
    &\textcolor{}{
    \hat {\mathcal{K}}^t_{u} \cup \check {\mathcal{K}}^{t-\tau_r-1}_{u} \cup \bar {\mathcal{K}}^{t-1}_u \setminus \tilde{\mathcal{K}}^t_{u}= (\bigcup_{v:(u,v)\in\mathcal{L}} \mathcal N_{uv}^t) \cup \bar {\mathcal{K}}^t_u}
    &\textcolor{}{\forall t\in \mathcal{T}\setminus\{0,\cdots,\tau_r\}, u\in\mathcal{U}}\label{vehicle_dynamic_t}\\ 
    &\check {\mathcal{K}}^t_{v} = \bigcup_{u\in\mathcal{U}:\tau_{uv} = \tilde t} \mathcal N^{t-\tilde t}_{uv}   &\forall t\in \mathcal{T},\tilde t\in \{0,1,\cdots,t\}, v\in\mathcal{U}\label{vehicle_dynamic_en_route}\\
    & \textcolor{}{
    \tilde{\mathcal{K}}^t_{u} = \hat {\mathcal{K}}^{t-\bar W}_{u}}
    &\textcolor{}{\forall  t\in \{\bar W,\cdots ,T-1\}, u\in\mathcal{U}} \label{vehicle_work_limit}
\end{align}

\noindent Based on the system dynamics elaborated above, we model this stochastic dynamic vehicle resource allocation problem as a Markov decision process (\citealp{Xu2018}; \citealp{Tang2020}). 

\noindent \textcolor{}{\textbf{Agents:} The fleet of available vehicles within a city is perceived as an agent. There exist $|\mathcal{U}|$ agents because the idle fleet within each city makes autonomous vehicle-line assignment decisions. The components of this MDP formulation $M\triangleq <\mathbf{S},\mathbf{A}, P, R>$ are defined as follows.}

\noindent \textbf{States:} The system state is captured by the tuple $\mathbf{S}= \{\mathbf{S}_h,\mathbf{S}_s,\mathbf{S}_d\}$. $\mathbf{S}_h$ indicates the current horizon of a day. $\mathbf{S}_s=\{\mathbf{S}_s^u\}_{u\in \mathcal{U}} = \{((\varphi_{(0)}^{uv})_v, \varphi_0^u, \varphi_1^u, \varphi_2^u,\cdots \varphi_{\tau_s}^u)\}_{u\in \mathcal{U}}$ describes the supply states of available vehicle resource in each city. In the first term, $\varphi_{(0)}^{uv}$ is the total number of empty seats in vehicles currently in service of line $(u,v)$, the vector $(\varphi_{(0)}^{uv})_v$ describes currently available seat resources in different lines originating from city $u$ without newly dispatching any vacant vehicle. $\varphi_{i}^u=\left| {\mathcal{K}}^{t+i}_u \right|$ is the accumulated number of vacant vehicles for assignment in the future $i$-th horizon. 
$\mathbf{S}_d = \{\mathbf{S}_d^{uv}\}_{(u,v)\in \mathcal{L}} = \{(\omega_{(0)}^{uv}, \omega_0^{uv}, \hat \omega_1^{uv}, \hat \omega_2^{uv},\cdots \hat \omega_{\tau_d}^{uv})\}_{(u,v)\in \mathcal{L}}$ describes the demand states in each line. $\omega_{(0)}^{uv}$ is the number of orders in line $(u,v)$ that will expire if not served within the current horizon. \textcolor{}{$\omega_{0}^{uv}$ is the number of orders in the matching pool when making the dispatching decisions, while $\hat \omega_{i}^{uv}$ is the estimated number of newly emerging orders in line $(u,v)$ in the future $i$-th horizon, which can be obtained from the historical average value at the same horizon. We set $\tau_s=3$ and $\tau_d = 5$ in subsequent numerical experiments in section \ref{sec_experiment}.}
The states not only specify the demand and supply information in the current horizon which will influence the instant reward but also describe the potential demand and supply dynamics in several future horizons, which is helpful to capture the spatial-temporal demand and supply fluctuation. Notice that more detailed attributes of supply and demand, such as the specific locations of vehicles and orders, are not included in $\mathbf{S}$, since we focus on approximating the effectiveness of resource allocation in the upper level of the framework and leave the evaluation of detailed information to the lower-level routing problem.

\noindent \textbf{Actions:} 
\textcolor{}{$\mathbf{A}=\{\mathbf{A}^u\}_{u\in \mathcal{U}}$ }represents the joint actions of all agents. 
In horizon $t$, the assignment decision of the idle fleet within city $u\in\mathcal{U}$ is a \textcolor{}{non-negative integer vector} $\mathbf{A}^u_t = ((\left | \mathcal N^t_{uv} \right|)_{v:(u,v)\in \mathcal{L}}, \left| \bar{\mathcal{K}}^t_u \right|)$, indicating the number of vehicles assigned to all lines originating from city $u$, and the number of vehicles reserved at the current city. \textcolor{}{Here, the action vector $\{\mathbf{A}^u_t\}_{u\in \mathcal{U}}$ has to satisfy the flow conservation constraints (\ref{vehicle_dynamic_0}) to (\ref{vehicle_work_limit}). The summation of the elements within vector $\mathbf{A}^u_t$ can not exceed the number of available vehicles within city $u$ at the beginning of this horizon, which is $|{\mathcal{K}}^{t}_u|=|\hat {\mathcal{K}}^t_{u} |+| \check {\mathcal{K}}^{t-\tau_r-1}_{u} |+| \bar {\mathcal{K}}^{t-1}_{u}|-| \tilde{\mathcal{K}}^t_{u}|$ and is denoted by $\varphi_0^u$ in the supply state tuple $\{\mathbf{S}_s^u\}_{u\in \mathcal{U}}$. The number of available vehicles $|{\mathcal{K}}^{t}_u|$ varies over horizons, which indicates that the feasible action space of each agent is state-dependent.}

\noindent \textbf{State transitions:} \textcolor{}{The state transition probability $\mathcal{P}(s'|s,a) = \mathcal{P}(\mathbf{S}_{t+1} = s'|\mathbf{S}_t = s, \mathbf{A}_t = a) $ explains the probability from state $s$ to next state $s'$ based on the selected action $a$. 
During one transition, the dispatching action $\{\mathbf{A}^u\}$ is executed to update the subsequent system states. The next horizon indicator $\mathbf{S}_h$ increases by one. The next supply state $\mathbf{S}_s$ is constructed following the constraints (\ref{vehicle_dynamic_0}) to (\ref{vehicle_work_limit}). The demand state $\mathbf{S}_d$ is updated by the matching results within the current horizon and the newly emerging orders. Since the supply state $\mathbf{S}_s$
consists of both the number of current idle vehicles ($\{(\varphi_{(0)}^{uv})_v\}$ and $\{\varphi_0^u\}$) and the number of vehicles that will be idle in future several horizons ($\{\varphi_i^u\}$), the distribution of the next supply states can be determined only based on current supply states and actions.
Thus, the fleet dispatching problem satisfies the Markov property under our definition on states and actions, which is $\mathcal{P}(\mathbf{S}_{t+1} = s_{t+1}|\mathbf{S}_0 = s_0,\mathbf{A}_0 = a_0,\cdots,\mathbf{S}_t = s_{t},\mathbf{A}_t = a_t) = \mathcal{P}(\mathbf{S}_{t+1} = s_{t+1}|\mathbf{S}_t = s_{t},\mathbf{A}_t = a_t)$.}

\noindent \textbf{Rewards:}
\textcolor{}{At dispatching horizon $t$, the platform determines the global dispatching decision $\mathbf{A}_t$ and receives an immediate reward $R_t=r(\mathbf{S}_t,\mathbf{A}_t)$ by assigning idle vehicles to intercity lines following the instruction of $\mathbf{A}_t$.
The immediate reward of the whole platform is the summation of the immediate reward of each agent respectively, i.e. $R_t = \sum_{u\in \mathcal{U}}R_t^u$. The immediate reward is defined as the total profits earned by vehicles dispatched at this horizon, minus the penalties due to lost orders during this horizon. }

\noindent \textcolor{}{To calculate the immediate rewards in the simulator environment and in practice, we need to solve the dynamic vehicle routing problems based on the system states and dispatching decisions $\mathbf{A}_t$. Let $R^{uv}$ denote the one-way trip fare of line $(u,v)\in \mathcal{L}$, $C$ denote the vehicle cost per distance, and $p_e$ denote the penalty rate for losing one passenger. Then, the rewards for the agent representing fleet within city $u$ at horizon $t$ are calculated as follows.
\begin{align}
R^u_t = \sum_{v:(u,v)\in \mathcal{L}}(\sum_{k\in\mathcal  N^t_{uv}} R^{uv}s_k^t - \sum_{k\in\mathcal  N^t_{uv}} C\delta_k^t-p_e R^{uv} \epsilon_{uv}^t)\label{r^u_t}
\end{align}
The first part of the formula corresponds to the whole-trip revenue of vehicles dispatched at horizon $t$, where $s_k^t$ is the number of passengers that vehicle $k$ picked in this trip. The second part corresponds to the traveling costs of the whole trip of all vehicles in $\mathcal  N^t_{uv}$, where $\delta_k^t$ is the total trip distance of vehicle $k$. The last part corresponds to the order loss penalty, where $\epsilon_{uv}^t$ is the number of lost passengers in line $(u,v)$ during horizon $t$. 
}

\noindent \textcolor{}{In the fleet dispatching problem, we seek to obtain an optimal vehicle allocation policy $\pi$, which is a mapping from states $\mathbf{S}$ and actions $\mathbf{A}$ to action probability distribution $\pi(\mathbf{A}|\mathbf{S})$. 
The objective of policy optimization is to maximize the expected system return, which is the expected discounted total rewards of all possible trajectories during the $T$ horizons conditional on the initial states $\mathbf{S}_0$. The overall objective function is shown as follows:
\begin{align}
\max_{\pi} \mathbb{E}_{\mathbf{A}\sim \pi(\cdot|\mathbf{S})}[ \sum_{t=0}^{T-1} \gamma^t r(\mathbf{S}_t,\mathbf{A}_t)|\mathbf{S}_0] \nonumber
\end{align}
where $\mathbf{A}\sim \pi(\cdot|\mathbf{S})$ represents the action $\mathbf{A}$ is obtained by sampling from a stochastic policy $\pi$ based on the state $\mathbf{S}$.}
For each horizon, the reward function $R_t = r(\mathbf{S}_t,\mathbf{A}_t)$ is determined by solving the lower-level problem based on the decision $\mathcal N_{uv}^t$, which makes it difficult to solve the dynamic resource allocation problem embedded with dynamic DARP using model-based approximate dynamic programming approaches. 
\textcolor{}{
Thus, we take advantage of reinforcement learning techniques to tackle the large-scale fleet dispatching problem. A more detailed description of our reinforcement learning model is provided in subsection
\ref{subsec_feudal}}.

\noindent \textcolor{}{Additionally, although we assume vehicles are identical in capacity and speed, they originate from various cities and commence work at different times, so vehicles are non-homogeneous in the rest of available work time at any given time and have different preferences on the final depots.
To focus on mitigating supply-demand imbalances, our MDP formulation is based on the quantities of vehicles in each area. Consequently, the direct decisions are the numbers of vehicles dispatched to each line. However, when assigning certain vehicles in practice, it is necessary to also consider their final depots. Therefore, we develop a mathematical programming-based approach to map quantity-based allocations into actual vehicle assignment schemes considering the heterogeneity of vehicles, which will be elaborated upon in subsection \ref{assignment}}.

\subsection{The vehicle routing model}\label{subsec_routing}
\noindent \textcolor{}{The lower level of the framework optimizes the multi-vehicle pooled-ride routing problem for each line separately at each matching interval. 
For each line, the solution of the dynamic DARP consists of the complete routes of all vehicles, including vehicles en route and newly dispatched  vehicles.
Each route originates from the current location of the vehicle and ends at the depot of the destination city.} 
Thus, routes planned at earlier matching intervals can be updated to serve newly emerging orders, but the vehicle-order matching obtained in earlier intervals cannot be violated. 
Vehicles travel along the currently planned routes at the current matching interval and update their routes based on new routing results at the beginning of the next interval. This process is repeated at each matching interval to enable online vehicle routing in operations. \par

\noindent We formulate the online routing problem of the line $(u,v)$ in a matching interval at the dispatching horizon $t\in\mathcal T$ as a mixed-integer linear program (MILP). Let $\mathcal{K}_{uv}$ denote the set of vehicles that are currently in a ride of this line, which also includes newly assigned vehicles $\mathcal N^{t}_{uv}$ if this is the first matching after vehicle dispatching by the upper-level problem. The order set $\mathcal{O}_{uv}$ includes all orders in the matching pool and the orders already matched by vehicles en route.

\noindent Let $v_k$ be the location of vehicle $k\in \mathcal{K}_{uv}$ at the beginning of this matching interval. \textcolor{}{For vehicle $k$, this trip will end at the depot $E$ in the destination city, and we set the latest arrival time $\check U_k$ based on its departure time for the current trip. Setting the latest time to arrive at the depot ensures that the total duration of the vehicle's route does not exceed a reasonable range, preventing the situation where a vehicle waits excessively for an order, which is crucial for improving the utilization of vehicle resources.} For order $p\in \mathcal{O}_{uv}$, the number of passengers is denoted as $n_p$, the origin is denoted as $s_p$ with time window $[\check L_p, \check U_p]$, the destination is denoted as $f_p$ with time window $[\hat L_p, \hat U_p]$. 
Notice that the vehicle-order matching obtained in earlier matching intervals cannot be violated. Thus, orders in $\mathcal{O}_{uv}$ can be classified into three categories. The first category $\mathcal{O}^1_{uv}$ are orders that have already been picked, and we use $k_p$ to denote the vehicle that has served order $p$. The second category $\mathcal{O}^2_{uv}$ are orders that have already been matched but not picked, and we use $k_p$ to denote the vehicle that has matched order $p$. The third category $\mathcal{O}^3_{uv}$ are orders still in the matching pool. Since orders in $\mathcal{O}^1_{uv} \cup \mathcal{O}^2_{uv}$ have been matched in feasible routing solutions of the last few intervals, there must exist a feasible solution to complete the service of these orders at the current interval. 

\noindent The routing problem is based on the complete graph $G(\mathcal V, \mathcal A)$. The vertices $\mathcal V= \mathcal V_k\cup \mathcal V_f\cup \mathcal V_s\cup \{E\}$, where $ \mathcal V_k=\{v_k,\forall k\in \mathcal{K}_{uv}\}$ indicates the locations of all vehicles at the beginning of the current interval, $\mathcal V_f=\{f_p,\forall p\in \mathcal{O}_{uv}\}$ indicates the destination nodes of all orders, $\mathcal V_s=\{s_p,\forall p\in \mathcal O^2_{uv}\cup \mathcal O^3_{uv} \}$ indicates the origin nodes of the unserved orders, and $E$ indicates the depot in the destination city. 
Let $R^{uv}$ be the one-way trip fare on line $(u,v)$, $D_{ij}$ be the distance of arc $(i,j)\in\mathcal A$, $C$ be the travel cost per distance. The objective of the routing problem is to maximize the revenue of all vehicles, subtracted by traveling costs from their current locations to the depot in city $v$. The objective function is shown as follows.
\begin{align}
    &\max\sum_{p\in \mathcal{O}_{uv}}\sum_{k \in \mathcal{K}_{uv}}R^{uv}n_{p}d_{pk}-C\sum_{(i,j)\in \mathcal A}\sum_{k\in \mathcal{K}_{uv}}y_{ij}^kD_{ij}\nonumber
\end{align}
Decision variables in the vehicle routing model are given by $x^{pk}_{ij}$, $y^{k}_{ij}$, $d_{pk}$ and $u_{kj}$. The first three kinds of variables are \textcolor{}{binary} variables, where $x^{pk}_{ij}$ indicates whether vehicle $k$ passes arc $(i,j)$ with passengers of order $p$, $y^{k}_{ij}$ indicates whether vehicle $k$ passes arc $(i,j)$, $d_{pk}$ indicates whether vehicle $k$ serves order $p$, $u_{kj}$ is a positive variable indicating the time for vehicle $k$ to arrive at node $j$. The constraints can be formulated as follows.
\begin{align}
    &\sum_{j\in \mathcal V\setminus \{i\}}\sum_{k\in \mathcal{K}_{uv}}y_{ij}^k \leq 1&\forall i \in \mathcal V \label{routing_flow_1}\\
    &\sum_{j\in \mathcal V\setminus \{i\}}y_{ji}^k - \sum_{j'\in \mathcal V\setminus \{i\}}y_{ij'}^k =0 & \forall i \in \mathcal V_f\cup \mathcal V_s, k\in \mathcal{K}_{uv} \label{routing_flow_2} \\
    &\sum_{j\in \mathcal V\setminus \{v_k\}}y_{v_kj}^k = \sum_{j\in \mathcal V\setminus \{E\}}y_{jE}^k = 1 & \forall k\in \mathcal{K}_{uv} \label{routing_flow_3}\\
    &d_{pk_p} = 1 & \forall p \in \mathcal O^1_{uv}\cup \mathcal O^2_{uv} \label{routing_match_1}\\
    &\sum_{k\in \mathcal{K}_{uv}} d_{pk} \leq 1\qquad & \forall p \in \mathcal O^3_{uv} \label{routing_match_2}\\
    &\sum_{j\in \mathcal V\setminus  \{f_p\}}x^{pk}_{jf_p}-\sum_{j\in \mathcal V\setminus  \{f_p\}}x^{pk}_{f_pj} =d_{pk} & \forall p \in \mathcal{O}_{uv}, k\in \mathcal{K}_{uv}  \label{routing_match_3}\\
    &\sum_{j\in \mathcal V\setminus  \{s_p\}}x^{pk}_{s_pj}-\sum_{j\in \mathcal V\setminus  \{s_p\}}x^{pk}_{js_p} =d_{pk} & \forall p \in\mathcal O^2_{uv}\cup \mathcal O^3_{uv}, k\in \mathcal{K}_{uv}  \label{routing_match_4}\\
    &\sum_{j\in \mathcal V\setminus \{i\}}x^{pk}_{ij} - \sum_{j'\in\mathcal V\setminus \{i\}}x^{pk}_{j'i} = 0 &\forall p \in \mathcal{O}_{uv},  i\in \mathcal V_f\cup \mathcal V_s\setminus \{s_p,f_p\}, k\in \mathcal{K}_{uv} \label{routing_match_5}\\
    & x^{pk}_{ij}\leq y^k_{ij} & \forall (i,j)\in \mathcal A, k\in \mathcal{K}_{uv} \label{routing_match_6}\\
    &\sum_{p\in \mathcal{O}_{uv}}x^{pk}_{ij}n_p\leq Wy_{ij}^k & \forall (i,j) \in \mathcal A,  k\in \mathcal{K}_{uv} \label{routing_cap_1}\\
    & u_{kv_k} = t_0 & \forall k\in \mathcal{K}_{uv}\label{routing_time_1}\\
    & u_{kj}-u_{ki} \geq D_{ij}/s-M(1-y_{ij}^k) & \forall (i,j) \in \mathcal A, k\in \mathcal{K}_{uv}\label{routing_time_2}\\
    & \check L_{p}d_{pk}\leq u_{ks_p} \leq \check U_{p}d_{pk} & \forall p \in \mathcal O^2_{uv}\cup \mathcal O^3_{uv},  k\in \mathcal{K}_{uv}  \label{routing_time_3}\\
    & \hat L_{p}d_{pk} \leq u_{kf_p} \leq \hat U_{p}d_{pk}& \forall p \in \mathcal{O}_{uv},  k\in \mathcal{K}_{uv}  \label{routing_time_4}\\
    & \textcolor{}{u_{kE} \leq \check U_{k}}& \textcolor{}{\forall   k\in \mathcal{K}_{uv}}  \label{routing_time_5}\\
    & x^{pk}_{ij} \in \{0,1\}, y^{k}_{ij} \in \{0,1\}, d_{pk} \in \{0,1\}& \forall p\in \mathcal{O}_{uv}, (i,j)\in \mathcal A ,k\in \mathcal{K}_{uv} \label{routing_decision_1}
\end{align}

\noindent Constraint (\ref{routing_flow_1}) ensures that each node can be passed at most once, and each arc is passed by at most one vehicle. If more than two vehicles pass the same arc, there must exist a shorter route for one of them. Constraints (\ref{routing_flow_2}) and (\ref{routing_flow_3}) are flow conservation constraints that the route of any vehicle in $\mathcal{K}_{uv} $ originates from its current location and ends the depot in city $v$. Constraints (\ref{routing_match_1}) and (\ref{routing_match_2}) indicate the matching between vehicles and orders. The matching of orders in $ \mathcal O^1_{uv}$ and $\mathcal O^2_{uv}$ are determined before this matching interval, while each order in $\mathcal O^3_{uv}$ can be matched to at most one vehicle en route. Constraints from (\ref{routing_match_3}) to (\ref{routing_match_6}) represent that passengers are picked by the matched vehicle at the origin node and delivered to the destination. Constraint (\ref{routing_cap_1}) ensures that the number of passengers in the vehicle will not exceed the capacity $W$. \textcolor{}{Constraints from (\ref{routing_time_1}) to (\ref{routing_time_5}) determine the time for vehicles to serve each node, where $t_0$ represents the current time, $s$ represents the vehicle speed, and $M$ is a very large positive value in constraint (\ref{routing_time_2}).} Based on the assumptions of hard time windows, vehicles have to serve the matched orders within the requested time windows. Vehicles may wait at the node if they arrive before the time window, while late arrivals lead to infeasible solutions.

\noindent \textcolor{}{The multi-vehicle pooled-ride routing problem is NP-hard because it contains the traveling salesman problem as a special case(\citealp{Ropke2006}).} Since we have to solve this problem repeatedly at every matching interval, we leverage the ALNS heuristic to obtain a near-optimal solution efficiently. The detailed heuristic algorithm is introduced in subsection \ref{subsec_darp}.  \par

\section{Solution Methodologies} \label{sec_methodology}
\noindent To tackle the challenges introduced in section \ref{sec_description}, we propose a two-level reinforcement learning framework to implement dynamic fleet dispatching and online vehicle routing on intercity ride-pooling services. A novel multi-agent hierarchical reinforcement learning method is adopted in the upper level of the framework to optimize the spatial-temporal allocation of vehicle resources by assigning idle vehicles to lines. The lower level of the framework uses the ALNS heuristic to solve the proposed online multi-vehicle pooled-ride routing problem efficiently and effectively based on the assignment decision of the upper level. 

\begin{figure}
  \centering
  \includegraphics[width=0.8\linewidth]{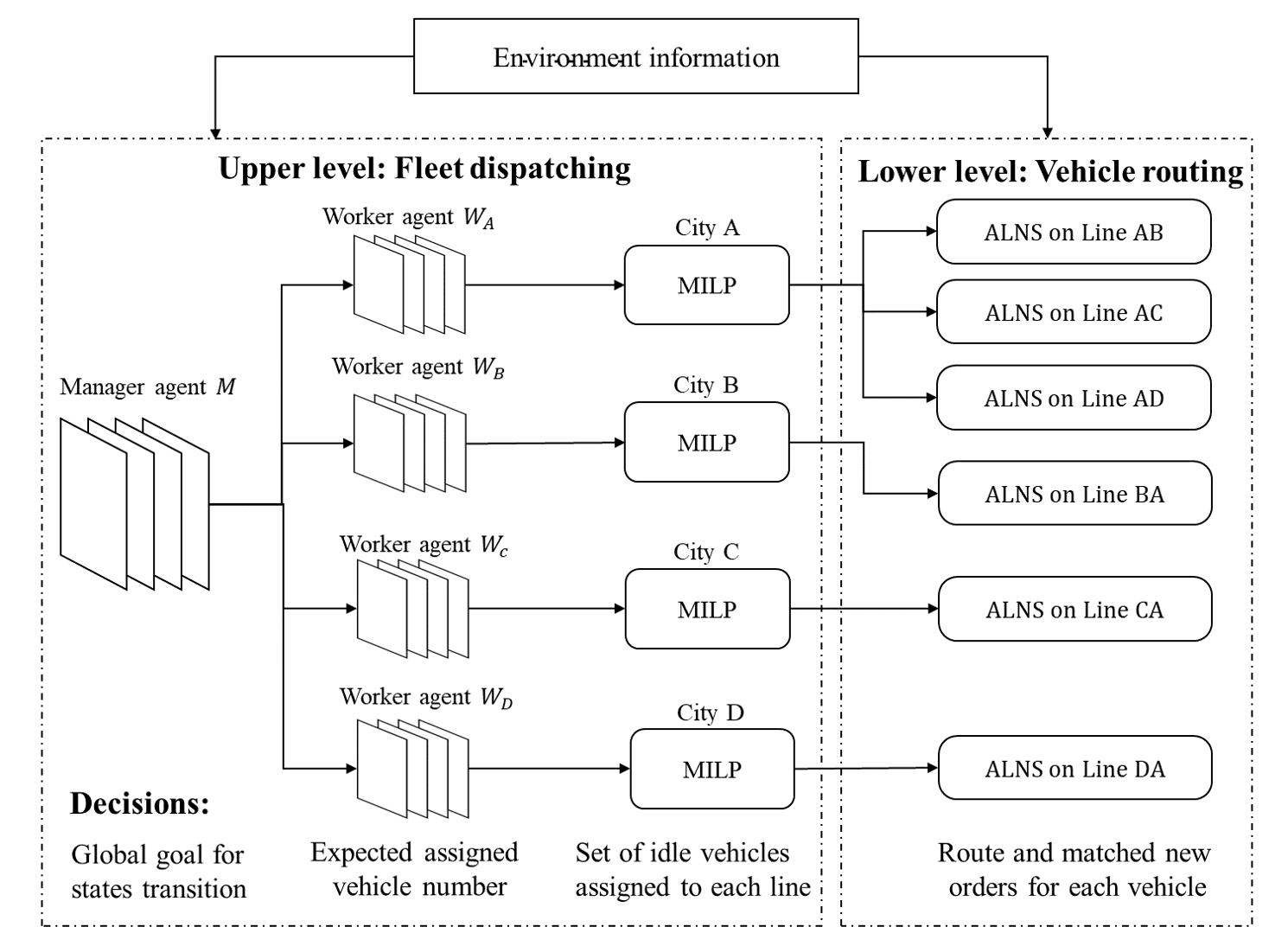}
  \caption{The fleet operational decision process for a network with one central city \emph{A} and three surrounding cities \emph{B}, \emph{C}, and \emph{D}.}
  \label{fig:exp_framework}
\end{figure}

\noindent Before delving into the specifics of the model, we provide an example that outlines the workflow across two levels of the framework. Consider the ride-pooling service provided in a city cluster with one central city \emph{A} and three surrounding cities \emph{B}, \emph{C}, and \emph{D}. Six intercity lines are available, including \emph{AB}, \emph{BA}, \emph{AC}, \emph{CA}, \emph{AD} and \emph{DA}. The corresponding operational decision process at the beginning of a dispatching horizon can be illustrated as shown in Figure \ref{fig:exp_framework}. Both levels of the framework observe distinct environmental information. \textcolor{}{The upper-level model employs a manager-worker hierarchical structure. The manager derives a global goal on state transitions. Each worker agent represents the available fleet within a city and determines the expected 
number of vehicles dispatched to each line based on the manager's instruction and local observations. A mixed integer linear program is executed independently for each city, mapping the assigned vehicle numbers to the specific vehicle assignments. Then the lower level of the framework utilizes the ALNS heuristic to address the routing problem for each line separately at each matching interval so that vehicles can serve the matched orders following the sequence planned}.

\subsection{Upper level: Multi-agent Feudal Networks} \label{subsec_feudal}
\noindent The upper level of the framework is expected to learn the optimal vehicle dispatching policy for each city within the city cluster. In our study, the fleet of idle vehicles within each city is regarded as an agent, multiple agents operate collaboratively to improve the system throughput. Thus, we use a multi-agent reinforcement learning framework at the upper level. MARL has been widely used in fleet operations in ride-sourcing platforms, which performs well in scalability in large-scale problems (\citealp{xu2018large}; \citealp{guo2020deep}; \citealp{jiao2021real}; \citealp{kullman2022dynamic}). However, MARL has a prominent drawback that the training process may be non-stationary when each agent updates its own policy. The system may fail to converge to the global optimum because each agent makes decisions based on its own utility function and local observation. Conventional reinforcement learning methods, such as Q-learning or policy gradient methods, usually exhibit high variance in multi-agent environments where coordination of agents is required (\citealp{lowe2017multi}).

\noindent To handle the problem aroused by decentralized training, we propose a Multi-agent Feudal Network (MFuN) framework to enhance dispatching decision coordination among agents. Feudal reinforcement learning (FRL) was first proposed by \citet{dayan1992feudal}. \citet{vezhnevets2017feudal} introduced Feudal Networks (FuN) to extend conventional FRL to the form of deep neural networks, which perform well in long timescale credit assignment, especially in environments with sparse reward signals. A manager module and a worker module are constructed in FuN, where the manager sets goals as directions for the actions of the worker. The goals can be interpreted as an expected objective for global state transition in future horizons. The worker receives intrinsic rewards for actions consistent with goals. Multi-agent settings are introduced by \citet{ahilan2019feudal} to set multiple agents as workers. The shared goals can be used to encourage the cooperation of simultaneously-acting agents. This hierarchical framework is highly compatible with the fleet dispatching problem within the city cluster. The vehicle assignment decisions in each city are relatively independent during the current horizon, but the significant influence of these actions on the supply states of other agents propagates throughout the city network in subsequent horizons, thereby influencing the overall system profit. Under the guidance of a global manager, all agents are expected to cooperate to improve the effectiveness of vehicle resource allocation to mitigate the imbalance between demand and supply. 

\subsubsection{Model structure} \label{structure}

\begin{figure}
  \centering
  \includegraphics[width=0.6\linewidth]{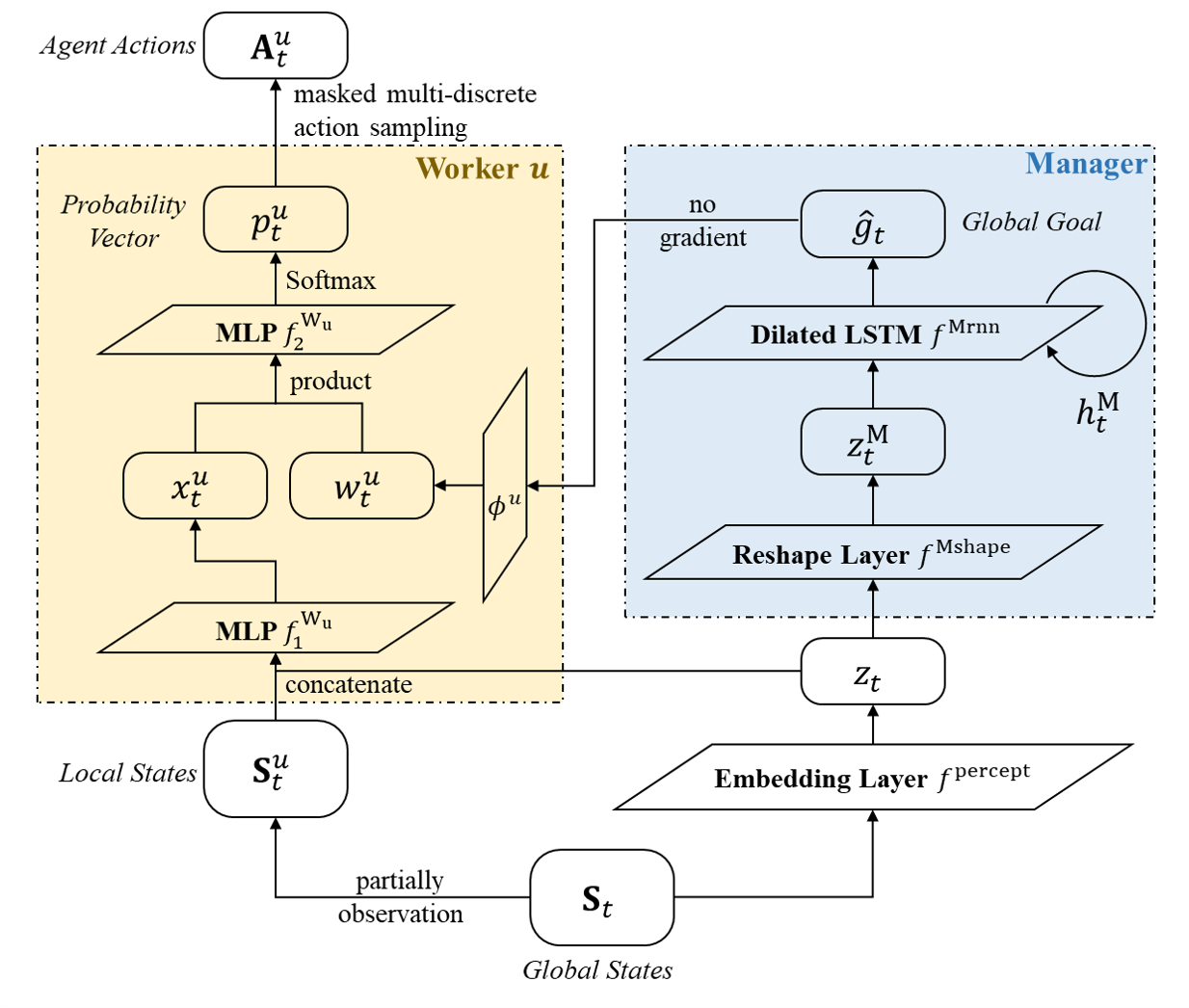}
  \caption{\textcolor{}{Actor network structure of MFuN}}
  \label{fig:model_structre}
\end{figure}

\noindent \textcolor{}{In this section, we introduce the model structure of MFuN, which maps states $\mathbf S_t= \{\mathbf S_{h,t}, \mathbf S_{s,t}, \mathbf S_{d,t}\}$ to actions $\mathbf A_t=\{((\left | \mathcal N^t_{uv} \right|)_{v:(u,v)\in \mathcal{L}}, \left| \bar{\mathcal{K}}^t_u \right|)\}_{u:u\in \mathcal U}$ as proposed in section \ref{sec_description}. The model structure of MFuN is shown as Figure \ref{fig:model_structre}}. The model contains a manager and several worker agents. The manager and worker agents share the embedding layer $f^{\mathrm{percept}}$ to obtain intermediate representation $z_t$ of global states $\mathbf S_t$. We use the same module for the manager $\mathrm M$ as proposed by \citet{vezhnevets2017feudal}, which contains a reshape layer $f^{\mathrm{Mshape}}$ and a recurrent neural network architecture $f^{\mathrm{Mrnn}}$. The output of the manager is a goal vector $\hat g_t$, which is obtained by a dilated LSTM model with input as reshaped $z_t$ and internal states $h^{\mathrm M}_t$. The dilated LSTM model is composed of $r$ separate groups of sub-states, where $r$ is a dilation radius. At \textcolor{}{dispatching horizon} $t$, only the $t\%r$ sub-states are updated, where $\%$ denotes the modulo operation, and the output is pooled across the previous $c$ outputs to accomplish smooth variation. By updating only part of the whole states at each time, the dilated LSTM model can preserve the memories for longer periods without missing any input experience. The forward dynamics of the \textcolor{}{manager} policy networks are given by Equations (\ref{manager1}) and (\ref{manager2}).
\begin{align}
& z^{\mathrm M}_t=f^{\mathrm{Mshape}}(z_t) \label{manager1}\\
& h^{\mathrm M}_t,\hat g_t =f^{\mathrm{Mrnn}}(z^{\mathrm M}_t,h^{\mathrm M}_{t-1})\label{manager2}
\end{align}
The module of a worker agent $\mathrm W_u$ is composed of two multi-layer perceptions (MLP) $f_1^{\mathrm W_u}$ and $f_2^{\mathrm W_u}$. Parameters in each worker module are not shared since they may differ in network scales. The mathematical formulations of the policy networks of worker agent $\mathrm{W}_u$ are given by Equations (\ref{worker1}), (\ref{worker2}), and (\ref{worker3}).
\begin{align}
& w^u_t = \phi^u(\sum_{i=t-c}^t g_i); g_t=\hat g_t/\Vert \hat g_t\Vert \label{worker1}\\
& x^u_t=f^{\mathrm W_u}_1(z_t\oplus\mathbf S^u_t)\label{worker2}\\
& \textcolor{}{p^u_t = \mathrm{Softmax}(f^{\mathrm W_u}_2(w^u_tx^u_t))}\label{worker3}
\end{align}

\noindent For each worker agent $\mathrm{W}_u$, the first MLP $f_1^{\mathrm{W}_u}$ takes the concatenation of global state representation $z_t$ and local observation $\mathbf S^u_t$ as input. Here, the local observation of agent $\mathrm W_u$ is composed of local supply and demand information. \textcolor{}{Similar to the definition in subsection \ref{dispatching}, the local supply states are $\mathbf{S}_s^u= ((\varphi_{(0)}^{uv})_{v:(u,v)\in \mathcal{L}}, \varphi_0^u, \varphi_1^u, \varphi_2^u,\cdots \varphi_{\tau_s}^u)$, including the number of currently available seat resources on the line originating from city $u$ and idle vehicle resources at future horizons. The local demand states are $\mathbf{S}_d^{u} =\{\mathbf{S}_d^{uv}\}_{v:(u,v)\in \mathcal{L}} = \{(\omega_{(0)}^{uv}, \omega_0^{uv}, \hat \omega_1^{uv}, \hat \omega_2^{uv},\cdots \hat \omega_{\tau_d}^{uv})\}_{v:(u,v)\in \mathcal{L}}$, which summarizes the number of various unserved orders in each line originating from city $u$. Each worker agent is also influenced by the current global goal $g_t$, which is a normalized vector of the output of the manager output $\hat g_t$. As shown in Equation(\ref{worker1}), the normalized system goal $g_t$ is then summed up with the last $c$ normalized goals, and projected by an agent-specified linear transformation $\phi^u$ into a local goal vector $w_t^u$, which reflects the comprehension of the global goals by worker agent $\mathrm{W}_u$ based on their own features}. 
\textcolor{}{The local goal vector $w_t^u$ is then combined with local intermediate representation $x^u_t$ via product to obtain logits vector $f^{\mathrm W_u}_2(w^u_tx^u_t)$. We decompose the logits vector into several parts to correspond to the multiple dimensions of actions. Each part comprises multiple real values representing the scores for different options within a specific action dimension. Subsequently, we apply the Softmax function separately to each part of the logits vector, thereby converting the real-valued scores over action options into normalized probability values for selecting the corresponding actions. Therefore, the probability vector $p_t^u$ also consists of multiple components, each representing a probability distribution over the action space within a specific dimension of actions. The values in each component ensure a sum of 1 respectively, indicating a valid probability distribution. Finally, we can sample each dimension of the action from $p_t^u$ independently, and then integrate each component to form the complete action vector $\mathbf{A_t^u}$.} 

\begin{figure}
  \centering
  \includegraphics[width=0.6\linewidth]{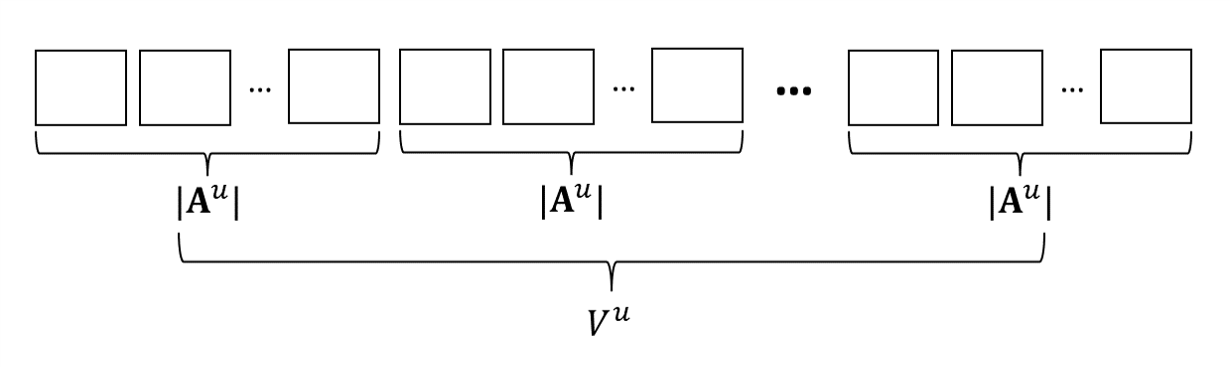}
  \caption{The output of worker agent $u$}
  \label{fig:action}
\end{figure}

\noindent \textcolor{}{So far, we have presented a thorough overview of the network structure. However, there remains a question of how to ensure the feasibility of the actions sampled. 
In the MDP formulation proposed in subsection \ref{dispatching}, the actions of the agent representing idle fleet within city $u$ are defined as $\mathbf{A}^u_t = ((\left | \mathcal N^t_{uv} \right|)_{v:(u,v)\in \mathcal{L}}, \left| \bar{\mathcal{K}}^t_u \right|)$, which is composed of $|\{v\}_{(u,v)\in \mathcal{L}}|+1$ dimensions. It is worth noting that the actions in different dimensions are not mutually independent. The summation of numbers of vehicles assigned to different lines can not exceed $ |{\mathcal{K}}^{t}_u|=|\hat {\mathcal{K}}^t_{u} |+| \check {\mathcal{K}}^{t-\tau_r-1}_{u} |+| \bar {\mathcal{K}}^{t-1}_{u}|-| \tilde{\mathcal{K}}^t_{u}|$, which is the number of all idle vehicles in city $u$ that are available for assignment. Thus, we can not use Softmax to directly determine the probability distribution over the number of vehicles dispatched to each line respectively. Instead, we use a virtual fleet of size $ V^u$ to represent all available vehicle resources and determine the dispatching action of each virtual vehicle, where $ V^u$ is a predetermined constant. 
Each virtual vehicle selects a destination city from $|\mathbf A^u|=|\{v\}_{(u,v)\in \mathcal{L}}|+1$ options, including city $u$ and its neighbors. Hence, as shown in Figure \ref{fig:action}, the output $p_t^u$ consists of $V^u$ components, each comprising a normalized probability distribution over $|\mathbf A^u|$ action options. When sampling for actions, we use decision masks to handle the situations of $ |{\mathcal{K}}^{t}_u|<V^u$, which represents the number of idle vehicles in reality is smaller than the predefined virtual fleet size. In such cases, the actions of the last $ V^u-|{\mathcal{K}}^{t}_u|$ virtual vehicles will be masked. Then, the sampled actions of the first $ \min\{|{\mathcal{K}}^{t}_u|,V_u\}$ virtual vehicles are recorded. By summing up the number of virtual vehicles dispatched to each line, we can clarify each element of $\mathbf{A}^u_t$. If the number of idle vehicles in reality exceeds $V^u$, the $ |{\mathcal{K}}^{t}_u|-V^u$ excess vehicles will be kept in the current city at the current horizon. The actions of different virtual vehicles can be treated as mutually independent, so our method can always guarantee the feasibility of the sampled actions.}

\noindent \textcolor{}{The selection of the quantity $ V^u$ is crucial for the model's performance. If $ V^u$ is so large that the number of available vehicles within city $u$ is always less than $ V^u$, then many components of the model outputs will always be blocked and unable to map to actions. This will significantly impact the convergence rate of the neural network. On the other hand, if $ V^u$ is so small that the number of available vehicles within the city often exceeds $ V^u$, then numerous idle vehicles cannot be timely dispatched, leading to a waste of vehicle resources. Thus, we usually set the value of $ V^u$ close to the median between the peak supply and the mean.}

\subsubsection{Learning methods}\label{learning}
\noindent We train the manager and workers with the Advantage Actor-Critic method, which learns the Actor parameters $\theta$ and the Critic parameters $\psi$ for all agents respectively. Many value-based learning methods, such as DQN and QMix, do not support multi-discrete action space, while the Advantage Actor-Critic method can still perform well. \textcolor{}{Notice that no gradients are propagated between workers and the manager, so the manager has to train its parameters autonomously to output latent goals for maximizing the system's expected return.
Both the manager and workers can evaluate the state value, output actions, receive extrinsic rewards, and learn from interacting with the environment. For each agent, the Actor is a policy function parameterized by $\theta$, which can provide the policy $\pi_\theta(\mathbf A|\mathbf S)$. The Critic is a value function $V_{\psi}(\mathbf S)$, which serves as a baseline to update the Actor's parameters. We have introduced the architecture of the Actor networks of the manager and workers in subsection \ref{structure}. Each Critic is established based on the corresponding Actor network but uses a different output layer to obtain single-valued state value estimation. The Critic of the manager supplements a linear layer $f^{\mathrm{Mlinear}}$ after the dilated LSTM to output the global state value $V_{\psi^{\mathrm M}}(\mathbf S_{t})$, i.e., $V_{\psi^{\mathrm M}}(\mathbf S_{t}) =f^{\mathrm{Mlinear}}(\hat g_t)$. The Critic of a worker $u$ use a linear layer $f^{\mathrm W_u}_{3}$ instead of the Softmax layer to obtain the local state value $V_{\psi^{\mathrm W}_u}(z^t,\mathbf S_{t})$, i.e., $V_{\psi^{\mathrm W}_u}(z^t,\mathbf S_{t}) =  f^{\mathrm W_u}_{3}(f^{\mathrm W_u}_2(w^u_tx^u_t))$. For each agent, the Actor and the Critic share most of the parameters within their corresponding networks. These parameters will be updated when minimizing the policy loss function and value loss function.
}

\noindent At each iteration, we conduct a simulation of the daily operations of the intercity ride-pooling services. The fleet dispatching decisions are made at each horizon to assign idle vehicles to lines in the simulation environment. After several iterations of interacting with the environment, we use batch gradient descent to update the policies of agents based on the transition experience of size $|\mathcal B|$. Let $\psi^{\mathrm M}$ and $\psi^{\mathrm W}_u$ be parameters in the Critic networks of manager and worker $u$,  $\theta^{\mathrm M}$ and $\theta^{\mathrm W}_u$ be parameters in Actor networks of manager and worker $u$. Then the policy loss function and value loss function of the manager are defined as follows:
\begin{align}
&A^{\mathrm M}_t = R_t+\gamma^{\mathrm M} V_{\psi^{\mathrm M}}(\mathbf S_{t+1}) -V_{\psi^{\mathrm M}}(\mathbf S_{t}) \label{manager_advantage}\\
& \mathcal{J}_{\psi^{\mathrm M}} = \frac{1}{2|\mathcal B|}\sum_{t\in\mathcal B} (A^{\mathrm M}_t)^2\label{manager_value_loss}\\
& \mathcal{J}_{\theta^{\mathrm M}} = -\frac{1}{|\mathcal B|}\sum_{t\in\mathcal B} A^{\mathrm M}_t {\mathrm d}_{\mathrm{cos}}(z^{\mathrm M}_{t+c}-z^{\mathrm M}_t,g_t(\theta^{\mathrm M}))\label{manager_policy_loss}
\end{align}

\noindent \textcolor{}{Equation (\ref{manager_advantage}) defines the advantage function of the manager, where $R_t$ is the immediate reward of the whole platform defined in subsection \ref{dispatching}, and $\gamma^{\mathrm M}$ is a discount factor for the manager.} 
We minimize the value loss function $\mathcal{J}_{\psi^{\mathrm M}}$ and policy loss function  $\mathcal{J}_{\theta^{\mathrm M}}$ to update parameters of the Critic and Actor networks. 
\textcolor{}{Here, we adopt the definition of the policy loss function proposed by \citet{vezhnevets2017feudal}. ${\mathrm d}_{\mathrm{cos}}(z^{\mathrm M}_{t+c}-z^{\mathrm M}_t,g_t(\theta^{\mathrm M}))$ is defined as the angle between actual state transitions in $c$ horizons and the normalized global goal $g_t$ to measure how system state transition follows the global goal, where ${\mathrm d}_{\mathrm{cos}}(X,Y) =X^TY/(|X||Y|)$ indicates the cosine similarity between two vectors. On the other hand, the system goal is compared with state transition in $c$ horizons, which provides a lower temporal resolution for the manager and enhances the foresight in our decision-making.}

\noindent The worker agents are trained in a similar manner, and the corresponding value loss function and policy loss function are shown as Equations (\ref{worker_intrinsic}), (\ref{worker_advantage}), (\ref{worker_value_loss}) and (\ref{worker_policy_loss}). The worker agent $u$ is not only driven by its local immediate reward $R^u_t$ but also encouraged by the intrinsic reward $R^I_t$ to follow the goals. \textcolor{}{The local immediate reward $R^u_t$ is defined in the formulation (\ref{r^u_t}). The intrinsic reward $R^u_t$ is the same for all workers, which is the average cosine similarities between system state transitions and the global goals in the past $c$ horizons. The manager leverages the intrinsic rewards to motivate workers to choose actions that can facilitate the development of the system in accordance with the planned goals. Since the manager works in a lower temporal resolution, the worker's policy with a higher degree of following the goals is more likely to avoid myopic vehicle dispatching decisions and keep valid in a large latent state space.}
\textcolor{}{The agents have to balance the trade-off between their rewards $R^u_t$ and the shared objective through a coefficient $\alpha$, which regulates the relative weight of intrinsic reward. We set $\alpha$ as 0.5 in the following numerical experiments.}
Then based on the experience trajectory, each worker agent can be trained separately via minimizing the value loss function $\mathcal{J}_{\psi^{\mathrm W}_u}$ and policy loss function  $\mathcal{J}_{\theta^{\mathrm W}_u}$.
\begin{align}
& R^{I}_t = \frac{1}{c}\sum_{i=1}^c {\mathrm d}_{\mathrm{cos}}(z^{\mathrm M}_t-z^{\mathrm M}_{t-i},g_{t-i})\label{worker_intrinsic}\\
& A^{{\mathrm W}_u}_t = R^u_t+\alpha R^{I}_t + \gamma^{\mathrm W} V_{\psi^{\mathrm W}_u}(z_{t+1},\mathbf S^u_{t+1})
-V_{\psi^{\mathrm W}_u}(z_{t},\mathbf S^u_{t})\label{worker_advantage}\\
& \mathcal{J}_{\psi^{\mathrm W}_u} = \frac{1}{2|\mathcal B|}\sum_{t\in\mathcal B} (A^{{\mathrm W}_u}_t)^2\label{worker_value_loss}\\
& \mathcal{J}_{\theta^{\mathrm W}_u} = -\frac{1}{|\mathcal B|}\sum_{t\in\mathcal B} A^{{\mathrm W}_u}_t \mathrm{log}(p^u_{t,\theta^{\mathrm W}_u}(z_{t},S^u_{t}))\label{worker_policy_loss}
\end{align}

\subsubsection{Determine vehicle assignment}\label{assignment}
\noindent \textcolor{}{Notice that the sampled actions only indicate the number of vehicles assigned to each line and reserved at the city, the dispatching decision for each vehicle has to be determined so that the assignment in reality is nearly consistent with the dispatched vehicle numbers in $\mathbf A_t$. A MILP is proposed to map the numbers in $\mathbf A_t$ into the sets of dispatched idle vehicles for each line respectively.} Let $\mathcal E_u$ be neighbors of the city $u$ (including the city itself $u$), and $a_v$ be the expected number of idle vehicles dispatched to the city $v\in \mathcal E_u$ obtained from $\mathbf A^u_t$. Among all idle vehicles in the city $u$ at dispatching horizon $t$, there may be some vehicles reaching maximum daily work time, which are expected to return to their original cities and get off work in time. Let $\mathcal F^t_u$ be the set of vehicles that will be off duty in a few horizons. The objective function of the MILP includes three parts as mathematical expressions (\ref{mapping_p1}), (\ref{mapping_p2}), and (\ref{mapping_p3}).
\begin{align}
&p_1=\sum_{v\in \mathcal E_u}\left |a_{v}-\sum_{k\in \mathcal K^t_u}x_{kv}\right |\label{mapping_p1}\\
&p_2=\sum_{v\in \mathcal E_u}\sum_{k\in \mathcal F^t_u}x_{kv}D_{vd_k}\label{mapping_p2}\\
&p_3=\sum_{k\in \mathcal F^t_u} \sum_{v\in \mathcal E_u\setminus \{u\}}x_{kv} \tilde r_{k}\label{mapping_p3}
\end{align}
where $x_{kv}$ is \textcolor{}{the binary decision variable} indicating whether vehicle $k$ is dispatched to city $v$ in this horizon. The mathematical expression (\ref{mapping_p1}) represents the gap between realistic assignment and sampled actions of worker agent $W_u$. The mathematical expression (\ref{mapping_p2}) represents the total distances for all vehicles in $\mathcal F^t_u$ to return origin cities after finishing the dispatched trip, where $D_{vd_k}$ indicates the distance between city $v$ and the origin city of vehicle $k$. \textcolor{}{Mathematical expression (\ref{mapping_p3}) is constructed from a fairness perspective, where $\tilde r_{k}$ is a predefined parameter representing the reward for dispatching vehicle $v$ to work rather than keep idling. The vehicles with lower percentages of time spent in transit compared with the average level are set $\tilde r_{k} = -1$, while others receive $\tilde r_{k} = 0$. Thus, we can minimize $p_3$ so that vehicles that spend more time in idle state are given higher priority to be assigned to lines rather than being reserved in the city.} Then the MILP can be formulated as follows.
\begin{align}
&\min \sum_{v\in \mathcal E_u}\beta_1 z_{v}+\beta_2p_2 +\beta_3p_3\nonumber\\
\text{s.t.} &\sum_{v\in \mathcal E_u}x_{kv}= 1 & \forall k\in \mathcal K^t_u \label{mapping_con1}\\
&\sum_{v\in \mathcal E_u}D_{vd_k}x_{kv}\leq D_{ud_k} & \forall k\in \mathcal F^t_u \label{mapping_con2}\\
&a_{v}-\sum_{k\in \mathcal K^t_u}x_{kv} \leq z_{v} & \forall v\in \mathcal E_u \label{mapping_con3}\\
&-a_{v}+\sum_{k\in \mathcal K^t_u}x_{kv} \leq z_{v} & \forall v\in \mathcal E_u \label{mapping_con4}\\
& x_{kv}\in \{0,1\}& \forall v \in \mathcal K^t_u,v\in \mathcal E_u 
\end{align}

\noindent The constraints include that each vehicle can only be dispatched to exactly one city, as shown in constraint (\ref{mapping_con1}), and that vehicle reaching maximum daily work time is not allowed to be dispatched to cities farther from its origin city, as shown in constraint (\ref{mapping_con2}). The objective function is a linear combination of mathematical expressions (\ref{mapping_p1}), (\ref{mapping_p2}) and (\ref{mapping_p3}). \textcolor{}{The weights $\beta_1$, $\beta_2$, and $\beta_3$ are set to firstly ensure that vehicles are dispatched consistently with $\mathbf A_t$, and consider other objectives afterward.} We linearize the absolute terms in (\ref{mapping_p1}) by introducing binary auxiliary variables $z_v$ to constraints (\ref{mapping_con3}) and (\ref{mapping_con4}). This MILP can be solved quickly using commercial software.

\noindent We conclude this subsection with the pseudo-code in Algorithm \ref{alg1}, \textcolor{}{which illustrates the complete process of training the proposed framework in a simulated environment.}

\begin{algorithm}[!ht]
    \renewcommand{\algorithmicrequire}{\textbf{Input:}}
    \renewcommand{\algorithmicensure}{\textbf{Output:}}
    \caption{Training process of MFuN}
    \label{alg1}
    \begin{algorithmic}[1] 
        \State Initialize parameters $\psi^{\mathrm M}$, $\{\psi^{\mathrm W}_u\}_u$, $\theta^{\mathrm M}$ and $\{\theta^{\mathrm W}_u\}_u$ ;
        
        \For {$i=1\to \mathrm{Iter}$}
            \State Reset the simulation environment, observe initial states $\mathbf S_0$;
            \For{$t=0\to T-1$}
                \State \textcolor{}{MFuN outputs $g_t$ and $\{p^u_t\}$ based on $\mathbf S_t$ and hidden states $h^{\mathrm M}_t$;}
                \State \textcolor{}{For each city, sample $\{\mathbf A^u_t\}$ based on $\{p^u_t\}$ };
                \State For each city, determine the assignment of each vehicle by solving the MILP proposed in subsection \ref{assignment};
                \State Solve the vehicle routing problem for each line at each matching interval;
                \State Record rewards $R_t$, $\{R^u_t\}$ and $R^I_t$, observe new environment states $\mathbf S_{t+1}$, reserve the transition experience;
            \EndFor
            \State Calculate advantages based on Equations (\ref{manager_advantage}), (\ref{worker_intrinsic}), and (\ref{worker_advantage});
            \State Update parameters based on Equations (\ref{manager_value_loss}), (\ref{manager_policy_loss}), (\ref{worker_value_loss}), and (\ref{worker_policy_loss}).
        \EndFor

 \end{algorithmic}
\end{algorithm}

\subsection{Lower level: Adaptive large neighborhood search heuristic}
\label{subsec_darp}
\textcolor{}{
\noindent At the lower level of the framework, we solve the dynamic multi-vehicle dial-a-ride problems proposed in subsection \ref{subsec_routing} using the ALNS heuristic at the beginning of each matching interval to generate a routing plan for each vehicle en route. The ALNS algorithm is executed for each line separately.
The inputs of the algorithm are the same as the basic settings in subsection \ref{subsec_setting}. The solution comprises the routes of all vehicles traveling in this line, with each route represented as a sequence of nodes for picking up and delivering passengers until reaching the depot}. During each iteration of the neighborhood search, a portion of orders in the current solution is removed by \emph{removal operators} while the \textcolor{}{nodes corresponding to some unfulfilled orders are inserted into certain positions of the sequence} by \emph{insertion operators} to construct a new solution. The choice of operators depends on the cumulative devotion to the objective function of each operator.

\noindent The neighborhood search process is initiated with a feasible solution. The initial routes are constructed based on the orders that have already been matched, following the sequence determined in earlier periods. \textcolor{}{The pick-up and delivery nodes of the} unmatched orders are then inserted into the routes using the \textbf{Regret-2 method}. This method identifies the optimal position for inserting each order into each route. This is achieved by recursively evaluating all feasible positions given the order and the route. For each order, the difference between the profit obtained from the best insertion position and the second-best insertion position is calculated as the Regret-2 value. At each iteration, the order with the highest Regret-2 value is inserted into its best position, and the Regret-2 values for the remaining unmatched orders are updated for the subsequent insertions. This process continues until all feasible insertions have been completed.

\noindent One basic operation in ALNS is to evaluate the insertion of a single order (both pickup and delivery nodes) into given feasible routes at the initialization stage and insertion stage. It is crucial to test the capacity feasibility and time window feasibility of an insertion, among which checking time window feasibility is extremely complex and time-consuming. \textcolor{}{We use the forward time slack method (\citealp{Gschwind2019}) to accelerate the check on whether the time window constraints of each node along the sequence are violated.}

\noindent The neighborhood search process is embedded in a simulated annealing metaheuristic. The initial temperature is $\tau_0$, which decreases at each iteration as $\tau\gets \tau\times c$, where $c$ is the cooling rate. The neighborhood search process ends at the iteration with a temperature lower than the stopping temperature $\tau_e$. ALNS learns over iterations which operators are more appropriate for the problem. Suppose there are $n$ removal operators $\mathcal R = \{r_1,r_2,r_3,...r_n\}$ and $m$ insertion operators $\mathcal I = \{i_1,i_2, i_3,... i_m\}$, where each operator $o$ is initialized with a score $s_o = 0$ and a weight $w_o = 1$. At each iteration, the operator is selected probabilistically based on the operator weights. For example, the probability of removal operator $r_i$ being selected is $P(r_i)=\frac{w_{r_i}}{\sum_{r_i\in \mathcal R} w_{r_i}}$. Let $F(\mathcal S)$ be the objective function value of the current solution $\mathcal S$, $F(\mathcal S')$ be the objective function value of the new solution $\mathcal S'$, $F(\mathcal S^{\mathrm{best}})$ be the objective function value of the historical best solution $\mathcal S^{\mathrm{best}}$. Suppose at a certain iteration, we choose operator $r_p\in \mathcal R$,$i_q \in \mathcal I$, then there exist four situations after the removal stage and insertion stage. The details of score updates are explained in pseudo-code in Algorithm \ref{alg2}. The parameters are set as $\theta_1>\theta_2>\theta_3>\theta_4$ such that operators leading to better solutions are enhanced adaptively in later stages. \textcolor{}{The weight of operator $o$ is updated as $w_o \gets 0.5 w_o+0.5\frac{s_o}{n_o}$, where $n_o$ is the number of times operator $o$ is used}. The operator selection criteria reflect the flexible balance between exploration and exploitation throughout the iterations.

\begin{algorithm}[!ht]
    \renewcommand{\algorithmicrequire}{\textbf{Input:}}
    \renewcommand{\algorithmicensure}{\textbf{Output:}}
    \caption{Adaptive large neighborhood search heuristic}
    \label{alg2}
    \begin{algorithmic}[1] 
        \Require Order information $\mathcal O_{uv}$, vehicle information $\mathcal K_{uv}$; 
        \Ensure Routing solution $\mathcal S$;
        \State Construct initial routing solution $\mathcal{S}$, $\mathcal S^{\mathrm{best}}\gets\mathcal{S}$;
        \State For each operator $o$, $s_o\gets 0$, $w_o\gets 1$;
        \State $\tau\gets\tau_0$;
        \While {$\tau>\tau_e$}
            \State For each operator $o$, $n_o\gets 0$;
            \For{$j=1\to \mathrm{Iter}$}
                \State Choose removal operator $r\in\mathcal R$ and insertion operator $i\in\mathcal I$ based on weights;
                \State $\mathcal S'\gets i(r(\mathcal S))$, $n_r\gets n_r+1$, $n_i\gets n_i+1$;
                \If {$F(\mathcal S')>F(\mathcal S^{\mathrm{best}})$}
                    \State $\mathcal S^{\mathrm{best}}\gets \mathcal S'$, $\mathcal S\gets \mathcal S'$, $s_r\gets s_r+\theta_1$, $s_i\gets s_i+\theta_1$;
                \ElsIf {$F(\mathcal S^{\mathrm{best}})\geq F(\mathcal S')>F(\mathcal S)$}
                    \State $\mathcal S\gets \mathcal S'$, $s_r\gets s_r+\theta_2$, $s_i\gets s_i+\theta_2$;
                \ElsIf {$ F(\mathcal S')\leq F(\mathcal S)$}
                    \State Generate random variable $\epsilon\in[0,1]$;
                    \If{$\epsilon\leq \mathrm{exp}(-\frac{F(\mathcal S)-F(\mathcal S')}{K\tau})$}
                        \State $\mathcal S\gets \mathcal S'$, $s_r\gets s_r+\theta_3$, $s_i\gets s_i+\theta_3$;
                    \Else
                        \State $s_r\gets s_r+\theta_4$, $s_i\gets s_i+\theta_4$;
                    \EndIf
                \EndIf
            \EndFor
            \State For each operator $o$, $w_o \gets 0.5w_o+0.5 \frac{s_o}{n_o}$;
            \State $\tau\gets \tau\times c$;
        \EndWhile
  
        \State \Return $\mathcal S^{\mathrm{best}}$.
 \end{algorithmic}
\end{algorithm}

\noindent The operators used in our study are similar to \citet{Ghilas2016}. The removal operators include: 

(1)\textbf{Random removal operator}: randomly removes $\phi_r$ orders from the solution.

(2)\textbf{Shaw removal operator}: randomly removes 1 order, then removes $\phi_r-1$ orders that are \textcolor{}{similar in locations and time window constraints}.

(3)\textbf{Worst removal operator}: removes $\phi_r$ orders with the highest travelling cost.

(4)\textbf{Time-based removal operator}: removes $\phi_r$ orders with the largest pick-up time delay.

The insertion operators include:

(1)\textbf{Greedy insertion operator}: iteratively inserts the order with the maximum additional objective function value to its best feasible position of the routes, until \textcolor{}{no feasible insertion with positive profit can be made}.

(2)\textbf{Distance greedy insertion operator}: iteratively inserts an order with the shortest additional distance to its best feasible position of the routes, until \textcolor{}{no feasible insertion with positive profit can be made}.

(3)\textbf{Regret-$m$ insertion operator}: similar to the \textbf{Regret-2 method} introduced for initialization above, but repeatedly insert orders with the largest differences of profit between the best insertion position and the $m$-best insertion position. \textcolor{}{The procedure ends when no feasible insertion with positive profit can be made}.


\section{Numerical Experiments} \label{sec_experiment}

\noindent In this section, we conduct numerical experiments to verify the effectiveness of our proposed framework under different supply and demand scenarios on three city networks. The first two series of experiments are conducted on toy networks and the third series of experiments are based on the realistic network and operational data in Xiamen and its surrounding cities. All computational experiments are implemented on the computation platform with Intel Core i5-12600KF CPU@3.70GHz, 16GB RAM and RTX3070 (8GB) GPU. The algorithms are coded with Python 3.6. The neural network is constructed and trained with Pytorch. The mixed integer linear programs are solved by Gurobi 10.0.1.

\noindent A simulator is established to approximate the dynamics of intercity ride-pooling services, including the generation of orders and the operations of the vehicle fleet. Each vehicle starts working at a predetermined time and location. The system initially assigns vehicles to lines and then updates the routes of all vehicles in transit. Once vehicle dispatching and routing are executed, the states are updated accordingly. This procedure continues until the end of the operational time to complete a daily simulation. During the training process of the proposed reinforcement learning model, a full simulation of daily operations constitutes an episode of trajectory data for policy learning. To alleviate the computational burden and prevent memory overflow, we simplify the training process in the simulation environment in three ways. \emph{Firstly}, during training iterations, the vehicle routing problems are only solved once in a dispatching horizon rather than solved at every matching interval. Vehicles travel along the planned routes during the horizon without intermediate route updating until the next assignment of idle vehicles. This simplification is reasonable when the newly emerging orders within a dispatching horizon are assumed to be known at the first matching process. \emph{Secondly}, \textcolor{}{compared to the settings in the literature focus on DARP (\citealp{Ghilas2016},\citealp{Gschwind2019}), fewer iterations are performed in ALNS} to solve the multi-vehicle routing problem in our numerical experiments. This approach \textcolor{}{significantly shortened} the training time since ALNS needs to be repeatedly solved in the simulation. \emph{Thirdly}, the vehicle routing problem is based on the Euclidean metric in both the toy network and the realistic network. We calibrate the detour ratio in the realistic network using trajectory data from operations to approximate the travel distances in reality.

\noindent \textcolor{}{Next, we introduce the basic settings of the simulation environment. We set the duration of a horizon to be 20 minutes for all subsequent experiments. In two toy networks, we consider daily intercity ride-pooling services for 40 horizons, while in the realistic network, we consider the services for 60 horizons (a daily simulation from 4:00 to 24:00). 
We set different fleet sizes for different experiments, and we will introduce the fleet size and their attendance settings in the corresponding subsections.
We consider seven-seater vehicles in all experiments, with each vehicle capable of serving up to six passengers. Drivers always have to take a rest of 20 minutes after an intercity trip for safety considerations, and the maximum working time for each driver is limited as $\bar W=10$ hours.
In experiments on the toy networks, the vehicle speed $s$ is set to 60 km/h, and the travel cost per kilometer $C$ is 1. In experiments on the realistic network, by calibrating based on the daily operation data, we set the average vehicle speed within the city to 25 km/h, the intercity average vehicle speed $s$ to 70 km/h, and the travel cost per kilometer $C$ to 0.5. }

\noindent \textcolor{}{The simulation environment can also generate demand as required. We generate order information for each line respectively based on specific order arrival rates. 
In experiments on toy networks, we sample the widths of the two time windows for each order from a normal distribution with a mean of 40 minutes and std of 15 minutes. And the duration from the latest pick-up time $\check U_p$ to the latest arrival time $\hat U_p$ and is sampled from a normal distribution with a mean of $2\times D_{uv}/s$ minutes and std of $0.25\times D_{uv}/s$ minutes, where $D_{uv}$ is the distance between two cities.
The coordinates of the origin and destination nodes of each order are uniformly distributed in their respective circular cities.
As for experiments on realistic networks, we will explain how to generate order information in subsection \ref{real}. 
The reservation lead time (the time from order placement to the beginning of the time window for picking-up) follows a truncated normal distribution $\mathcal{N}(40,30)$ within a range from 0 to 120 minutes. And the penalty rate for losing one passenger $p_e$ is set to 0.5. Other parameters related to network structures, such as the trip fare for each line, will also be provided in corresponding subsections.
}

\begin{figure}
  \centering
  \includegraphics[width=0.8\linewidth]{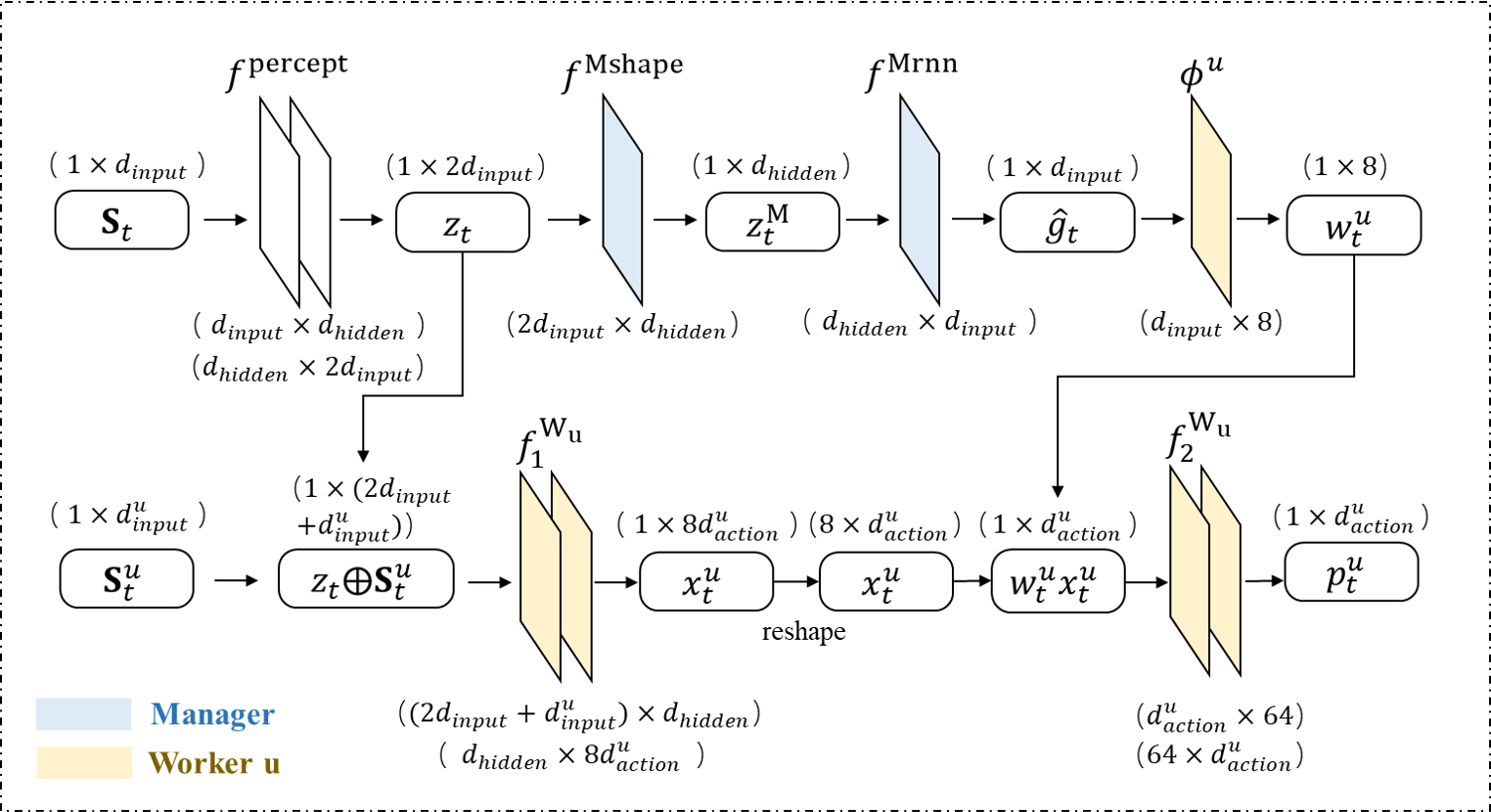}
  \caption{\textcolor{}{Neural network hidden layer size in MFuN}}
  \label{fig:nn_size}
\end{figure}

\noindent \textcolor{}{Our framework integrates the reinforcement learning method, MILP, and heuristic algorithm, leading to numerous hyperparameters for fine-tuning. We conducted several numerical experiments to identify good parameter combinations. Here, we present the values of the essential algorithm parameters employed in the subsequent experiments.
We first focus on the training procedure of the reinforcement learning method at the upper level of the framework. In each subsequent experiment, the neural networks are trained for 3000 episodes, and the test results come from trajectories of 20 episodes based on the well-trained neural networks. 
In MFuN, we set $\gamma^M = 0.99$ for the manager agent and $\gamma^W = 0.95$ for worker agents. For the manager, the number of looking forward horizons is set to $c=3$, and the dilation radius for manager LSTM is set to $r=5$.
We use RMSprop optimizer for manager and workers with a learning rate of the policy optimizer as $2.5\times 10^{-4}$. 
We use batch gradient descent to update the policies of agents and the batch size $|\mathcal B|$ is set to be 5 episodes.
In MFuN, we use the neural networks with hidden layer size shown as Figure \ref{fig:nn_size}. $d_{input}$ is the length of $\mathbf S_t$ and $d_{action}^u = V^u\times |\mathbf A^u|$ is the length of the output probability vector. The size of hidden layers in the neural networks scales up as the number of cities and lines increases. We set $d_{hidden}$ to 128 in experiments on toy networks and 256 in experiments on realistic networks.}

\noindent \textcolor{}{There also exists hyper-parameters for the MILP in subsection \ref{assignment}. We set the weights $\beta_1, \beta_2,\beta_3$ in the objective function to 5, 0.02, and 1 respectively. This parameter setting considers the distinct scales of each component in the objective function, to firstly ensure that vehicles are dispatched consistently with $\mathbf{A}_t$, and consider other objectives
afterward. As for control parameters in the ALNS, we set the hyper-parameters to balance the trade-off between computational time and the algorithmic performance.
For the outer loop, we set initial temperature $\tau_0=100$, stopping temperature $\tau_e=5$, and cooling rate $c=0.2$. 
And we set the number of iterations in the inner loop to 5, indicating the operator weights are updated every five times of neighborhood search.
The parameters for updating operator scores are set as $\theta_1=20$, $\theta_2=12$, $\theta_3=6$, $\theta_4=2$. At each neighborhood search, $\phi_r$, the number of orders removed, is set as one-fourth of the total number of unserved orders in the current solution.}

\noindent \textcolor{}{To better evaluate the effectiveness of the proposed MFuN framework on vehicle resource allocation among cities, we introduce a Myopic vehicle dispatching strategy as the benchmark. 
At each horizon, the Myopic approach determines the idle vehicle assignment for each city respectively. For each city, the Myopic method first estimates the number of vehicles required for serving currently unmatched orders on each line separately. If the total number of idle vehicles within a city is sufficient to fulfill the assignment, vehicles are randomly assigned to lines such that the number of vehicles dispatched to each line is nearly consistent with the planned number, while the remaining vehicles are reserved for the next dispatching. If there are not enough idle vehicles for this assignment, numbers of vehicles dispatched to each line are proportionally reduced until all idle vehicles are deployed. The Myopic method also uses ALNS to solve the vehicle routing problem at the lower level of the framework. 
Myopic method offers clear insights into the performance of decision-making based on the current states and exhibits stronger interpretability in contrast to RL methods. Thus we compare the optimization performance of the proposed MFuN framework with the Myopic method in subsections \ref{toy1}, \ref{toy2}, and \ref{real}.}
We also propose several reinforcement learning methods as benchmarks to show the advantage of MFuN in policy training. The details of these RL methods and their performances are presented in subsection \ref{train_method}. 
.

\noindent The evaluation of the proposed methods will be based on several key indices, including the average daily profit and the order fulfillment ratio. In the context of intercity passenger transportation, the platforms also emphasize the vehicle utilization rate, which represents the percentage of time vehicles spend in transit. In the subsequent sections, we will compare the performance of different methods based on these indices. This comparative analysis aims to evaluate the system's overall performance and illustrate the specific characteristics and properties of each method.

\subsection{Comparative analysis of network architectures}\label{train_method}

\noindent We first focus on the training process of the reinforcement learning method at the upper level of the framework. \textcolor{}{To validate the learning effectiveness of the MFuN network structure, we use three baselines with different network architectures to conduct a comparative analysis. }
\begin{itemize}
    \item \textcolor{}{\textbf{FuN}: The first baseline is the centralized decision feudal networks (FuN), and we use the Advantage Actor-Critic algorithm to train the networks. This model leverages a similar network architecture as MFuN but only consists of a single manager and a single worker. The worker networks jointly generate the dispatching decisions for all cities, so its outputs can be interpreted as the concatenation of the outputs of all workers in MFuN. }
    \item \textcolor{}{\textbf{MNFuN}: The second baseline is the multi-agent `non-feudal' feudal networks (MNFuN). MNFuN has the same network architecture as MFuN. The `non-feudal' represents that no intrinsic reward is used and the manager output $g_t$ is trained with gradients coming directly from the worker. In this framework, there is no clear boundary between the manager and the workers. The manager can only be regarded as the predecessor network of the worker because it cannot receive learning signals from the environment and get trained alone. We also use the Advantage Actor-Critic algorithm to train the MNFuN.}
    \item \textcolor{}{\textbf{MLP}: The third baseline uses a multi-layer perceptron (MLP) as the network and trains the model by the Proximal Policy Optimization algorithm (PPO). In the experiments on toy networks, we use an MLP that consists of three hidden layers of size 128, while in experiments on realistic networks, the hidden layer size is set as 256. Similar to FuN, this model also makes centralized decisions in a multi-discrete manner.}
\end{itemize}  
\textcolor{}{The outputs of different neural networks are then used to generate the agent actions by sampling and drive the simulation environment in the same way as the MFuN framework. By comparing the network architecture of MFuN with these baselines, we aim to assess its advantages in learning an intelligent policy in the stochastic dynamic vehicle resource allocation problem.}

\begin{figure}[!ht]
  \centering
  \includegraphics[width=1.0\linewidth]{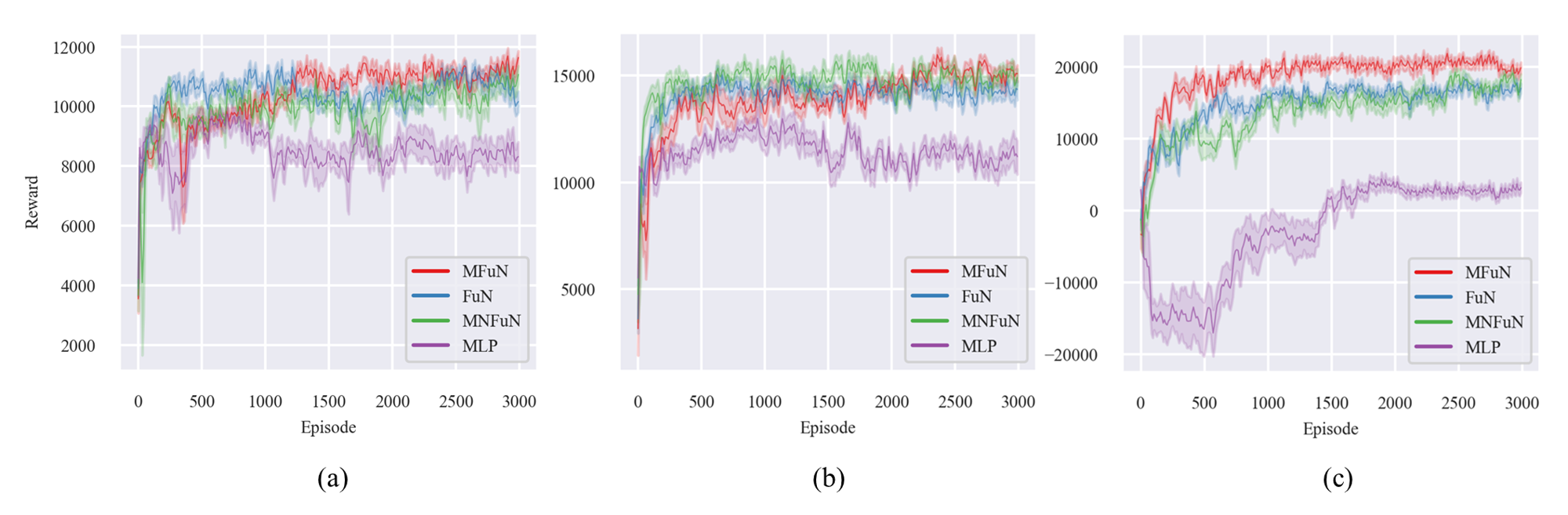}
  \caption{\textcolor{}{Learning curves of different RL methods in different networks. (a) The 2-city-2-line toy network. (b) The 3-city-4-line toy network. (c) Realistic network.}}
  \label{fig:lc_all}
\end{figure}

\noindent Figure \ref{fig:lc_all}(a) and (b) show the learning curves of the models on relatively small networks, which contain only 2 and 4 lines respectively. \textcolor{}{The demand and supply pattern on two toy networks will be introduced in subsequent subsections. The intercity distance is set as 60 km, and the trip fare is also set as 60. The red curves in the figures represent the reward obtained by the MFuN method over episodes, and green, blue, and purple curves represent three baselines respectively. 
Benefiting from the guidance of hierarchical reinforcement learning, the training results of MFuN, FuN, and MNFuN are significantly more stable than MLP. The results of MFuN and FuN are basically similar in this example, however, the advantage of multi-agent decision becomes more obvious when the network complexity increases. 
Figure \ref{fig:lc_all}(c) shows the learning curve on a realistic network with 7 cities and 12 lines. 
Compared with smaller networks, the advantage of hierarchical network architecture over fully connected networks is more pronounced. The learning curves of MLP exhibit dramatic fluctuations, while the training process of the hierarchical framework is more stable. 
Additionally, MFuN also achieves a higher daily profit compared to FuN in the task environment with large-scale networks.
Centralized decision approaches tend to exhibit relatively poor performance when the joint action space of agents expands exponentially.
This highlights the advantage of MFuN in scalability because each worker agent can learn its policy autonomously under the guidance of global goals, allowing for better performance as the number of agents grows. 
In networks with more cities and intercity lines, MFuN notably ensures effective cooperation among agents to achieve a pronounced advantage in the global return.}

\begin{figure}[!ht]
  \centering
  \includegraphics[width=1.0\linewidth]{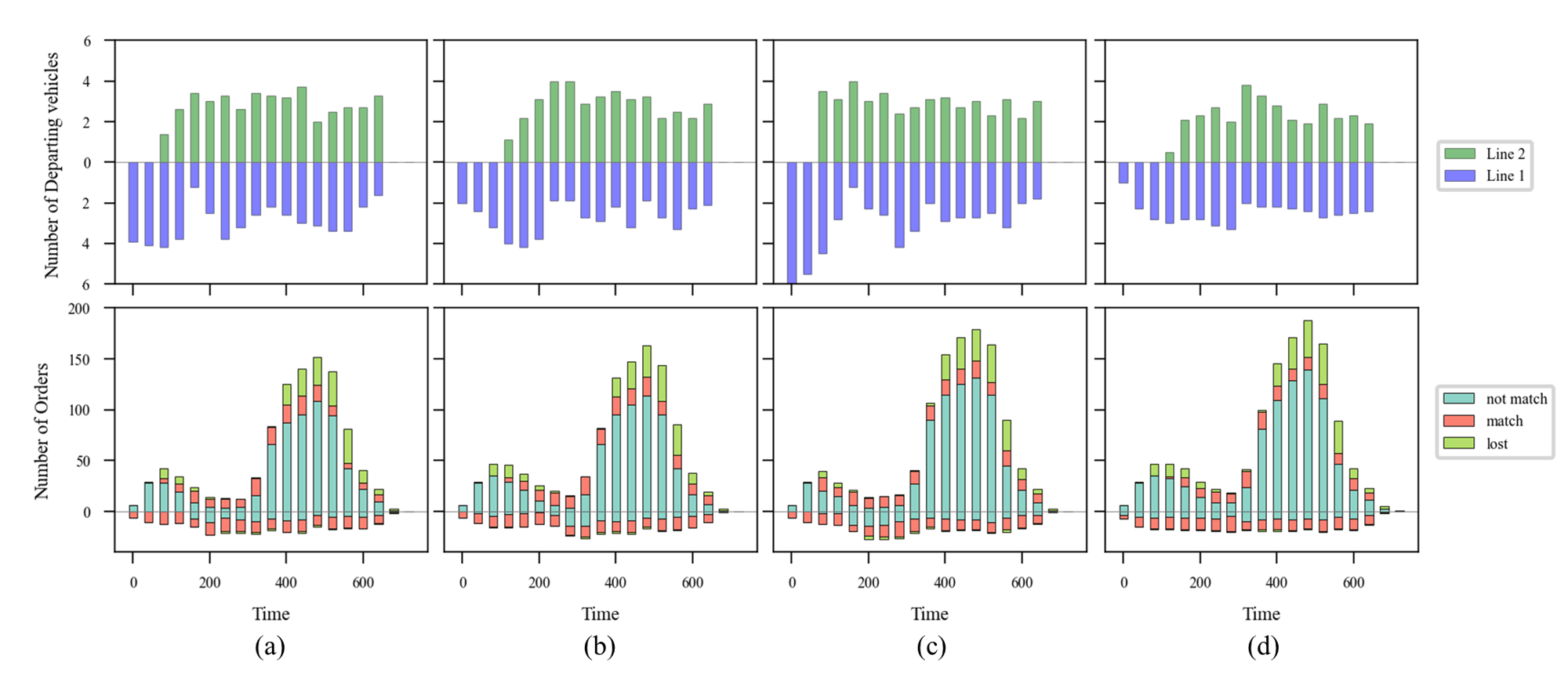}
  \caption{\textcolor{}{Number of departures and orders in each period. (a) MFuN. (b) MNFuN. (c) FuN. (d) Myopic.}}
  \label{fig:explain_learning}
\end{figure}

\noindent \textcolor{}{The comparison with baselines also highlights the advantage of MFuN in associating the actions of agents with subsequent extrinsic rewards and state transitions. We can gain insights into the policies learned by different models from observing the vehicle dispatching actions in the simulation environment on a 2-city-2-line network. Two intercity lines (line 1 and line 2) are operated between cities A and B. Line 1 (from A to B) experiences a steady low demand flow throughout the day while line 2 (from B to A) exhibits a demand peak as shown in Figure \ref{fig:toy1}. A fleet of 20 vehicles originating from city A is available for the intercity ride-pooling services of line 1 and line 2. A more detailed introduction to the demand pattern of this toy network will be provided in the next subsection. Figure \ref{fig:explain_learning} illustrates the vehicle dispatching patterns under the MFuN, MNFuN, FuN, and Myopic methods respectively. 
The upper figures show the average number of departures of idle vehicles at both lines within each two horizons (40 minutes), and the lower figures show the average number of lost orders (green), matched orders (red), and the orders not matched in this period but still available (cyan). 
Compared to the Myopic approach, all three reinforcement learning methods dispatch more vehicles from city A to city B, even when both lines are experiencing low demand. This proactive dispatching strategy ensures more sufficient vehicle resources available in city B ahead of the demand peak to reduce order loss. Among three reinforcement learning approaches, MFuN dispatches the most vehicles to line 2 during peak hours, resulting in the highest daily profits. The abundant vehicle supply of city B from 320 minutes to 440 minutes under MFuN's dispatching policy can be attributed to its accurate identification of periods with higher profits, so more vehicles are prepared at city B in advance. This underscores MFuN's heightened capability to associate the actions of each agent at each period with the expected global return in a long-term dynamic environment.}

\subsection{A 2-city-2-line toy network} \label{toy1}

\begin{figure}[!ht]
  \centering
  \includegraphics[width=1.0\linewidth]{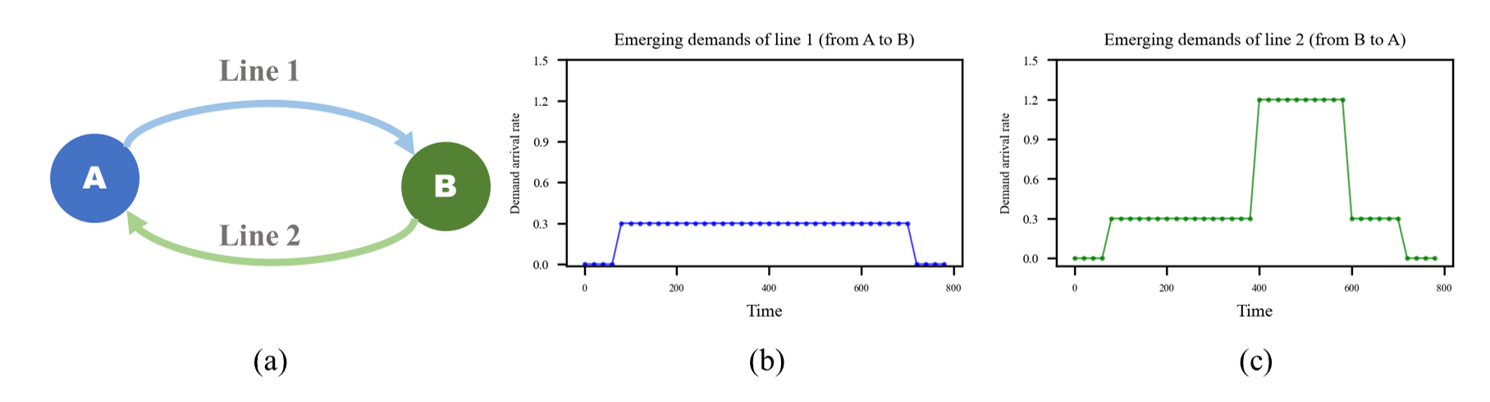}
  \caption{A 2-city-2-line toy network. (a) Topology of the network. (b) Demand from city A to B. (c) Demand from city B to A.}
  \label{fig:toy1}
\end{figure}

\noindent The 2-city-2-line toy network is the simplest and most basic intercity transportation network. Numerical experiments based on this network can demonstrate the effectiveness of the proposed framework on temporal scheduling. The dispatching decisions in the 2-city-2-line toy network illustrate how various methods adapt to fluctuations in supply and demand by only adjusting the departure time of each vehicle. Figure \ref{fig:toy1}(a) depicts the network topology of two cities, which have a radius of 15 km and the distance between two cities is 60 km. Figure \ref{fig:toy1}(b) portrays the demand of two lines. As introduced before, there is a relatively low demand for line 1 throughout the day, while there is a demand peak lasting for 200 minutes on line 2, which is four times higher than the demand during low-demand periods. \textcolor{}{Notice that the demand arrival rate here indeed represents the arrival rate for the beginning of the pick-up time windows, so the platform will receive the reservation earlier due to the reservation lead time.} 
In this series of experiments, a fleet of 20 homogeneous seven-seater vehicles is utilized. The vehicles start their operations in city A at the initial moment and are available to serve ride requests throughout the day.

\subsubsection{Sensitivity analysis of trip fare} \label{p_toy1}

\begin{figure}[!ht]
  \centering
  \includegraphics[width=1\linewidth]{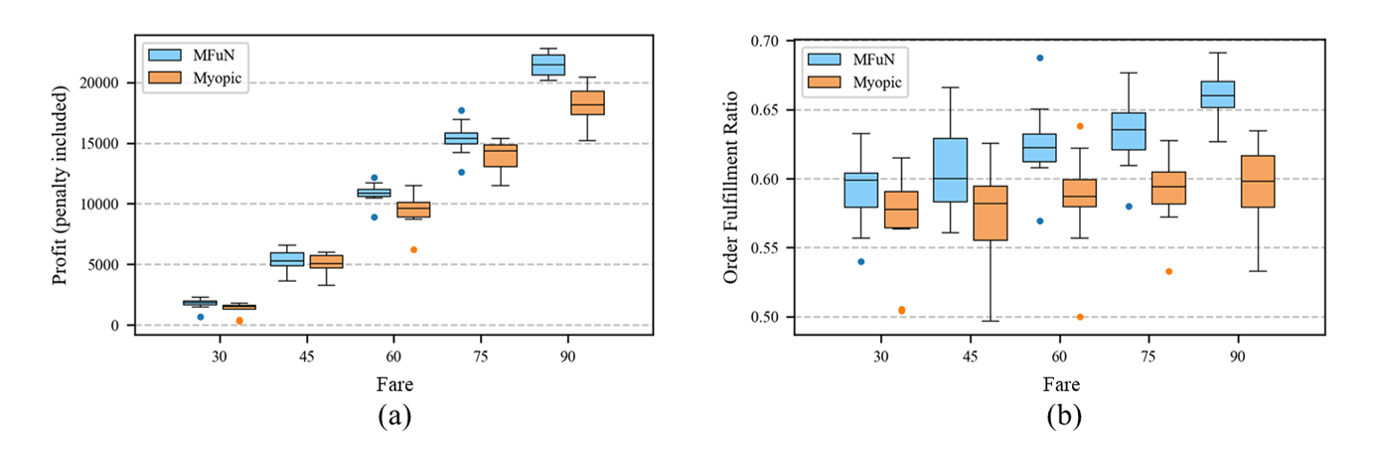}
  \caption{Performance of MFuN and Myopic in the 2-city-2-line toy network under different trip fares. (a) Profit. (b) Order fulfillment ratio.}
  \label{fig:p_toy1}
\end{figure}

\noindent In this subsection, we examine the impact of the one-way pooled-ride trip fare on the vehicle dispatching strategy. We assume equal trip fares for both directions, denoted as $R^{AB}=R^{BA}$. 
The trip fare values are varied, specifically set at 30, 45, 60, 75, and 90 for each passenger.
Given that the one-way traveling cost is at least 60, a vehicle must transport at least two passengers to attain profitability when the trip fare is relatively low. However, as the trip fare increases, the condition for vehicle profitability with respect to the occupancy rate becomes less stringent, allowing for more flexible dispatching strategies.

\noindent As depicted in Figure \ref{fig:p_toy1}, it can be observed that the Myopic method exhibits little sensitivity to changes in trip fare. Under this method, the one-way trip fare does not directly influence the dispatching schedule of vehicles. However, it does impact vehicle routing, as higher trip fares encourage picking up more passengers with longer detour distances within the city. Consequently, this leads to a slight increase in the order fulfillment ratio. In contrast, the MFuN framework demonstrates more intelligent vehicle scheduling. It consistently achieves higher daily profits and order fulfillment ratios compared to the Myopic method across various trip fare settings. Additionally, the increase in trip fare has a more pronounced positive impact on MFuN, particularly in improving the order fulfillment ratio. Under the high trip fare scenarios, MFuN achieves a significant improvement of 16.8$\%$ in the order fulfillment ratio during demand peak periods, and a 7.1$\%$ improvement in the daily average order fulfillment ratio, compared to the Myopic method.

\subsubsection{Sensitivity analysis of the distances between cities} 
\label{d_toy1}

\begin{figure}[!ht]
  \centering
  \includegraphics[width=1\linewidth]{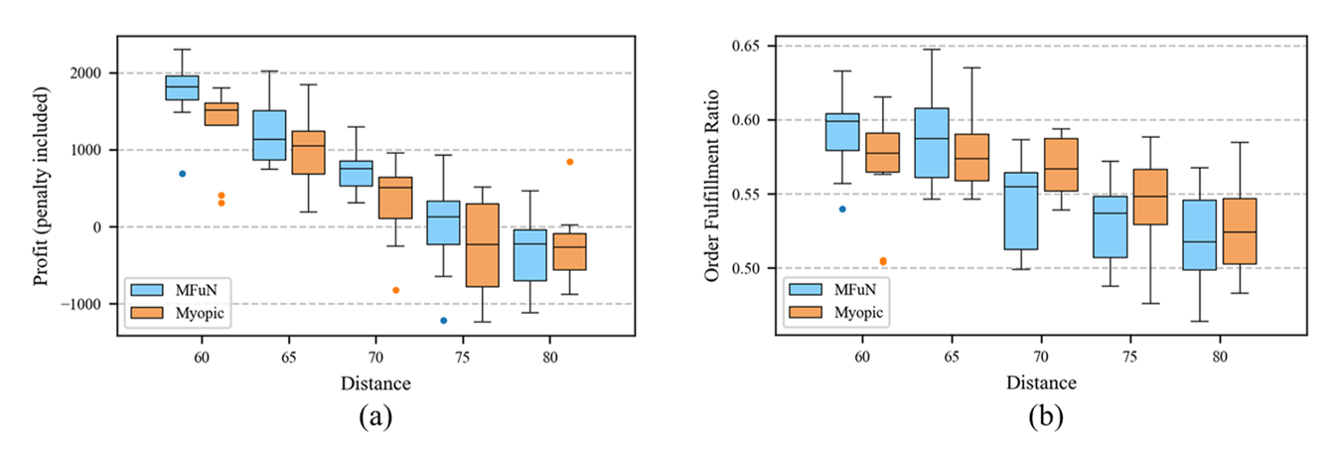}
  \caption{Performance of MFuN and Myopic in the 2-city-2-line toy network under different inter-city distances. (a) Profit. (b) Order fulfillment ratio.}
  \label{fig:d_toy1}
\end{figure}

\noindent In this subsection, we focus on the differences in dispatching strategies under different inter-city distances, ranging from 60 to 80, while keeping the one-way trip fare fixed at 30. Considering a one-way trip fare of 30 and an inter-city distance of 60 to 80, a vehicle must carry at least 2 to 3 passengers on a one-way trip to attain profitability. As depicted in Figure \ref{fig:d_toy1}, both the average daily profit and the order fulfillment ratio gradually decline as the distance between cities increases, regardless of the method employed. However, the MFuN approach demonstrates advantages in terms of profit in all scenarios, although it exhibits a lower order fulfillment ratio in cases where the distance is relatively large. The MFuN framework aims to enhance overall system profit by increasing the average number of passengers per trip, even if it means some orders may not be served due to a more conservative dispatching strategy. Furthermore, when the distance between cities is large, MFuN optimally utilizes the abundant demand during peak periods. This accumulation of vehicle resources in advance allows for a higher occupancy rate during peak periods, leading to improved efficiency. In contrast to subsection \ref{p_toy1}, where MFuN intensifies the frequency of departures to serve more orders in scenarios with high trip fares, in cases with a relatively large inter-city distance, MFuN reduces the frequency of departures to enhance one-way profitability. This showcases the MFuN framework's ability to propose diverse and efficient dispatching strategies tailored to different demand and supply scenarios.

\subsection{A 3-city-4-line toy network} \label{toy2}

\begin{figure}[!ht]
  \centering
  \includegraphics[width=1.0\linewidth]{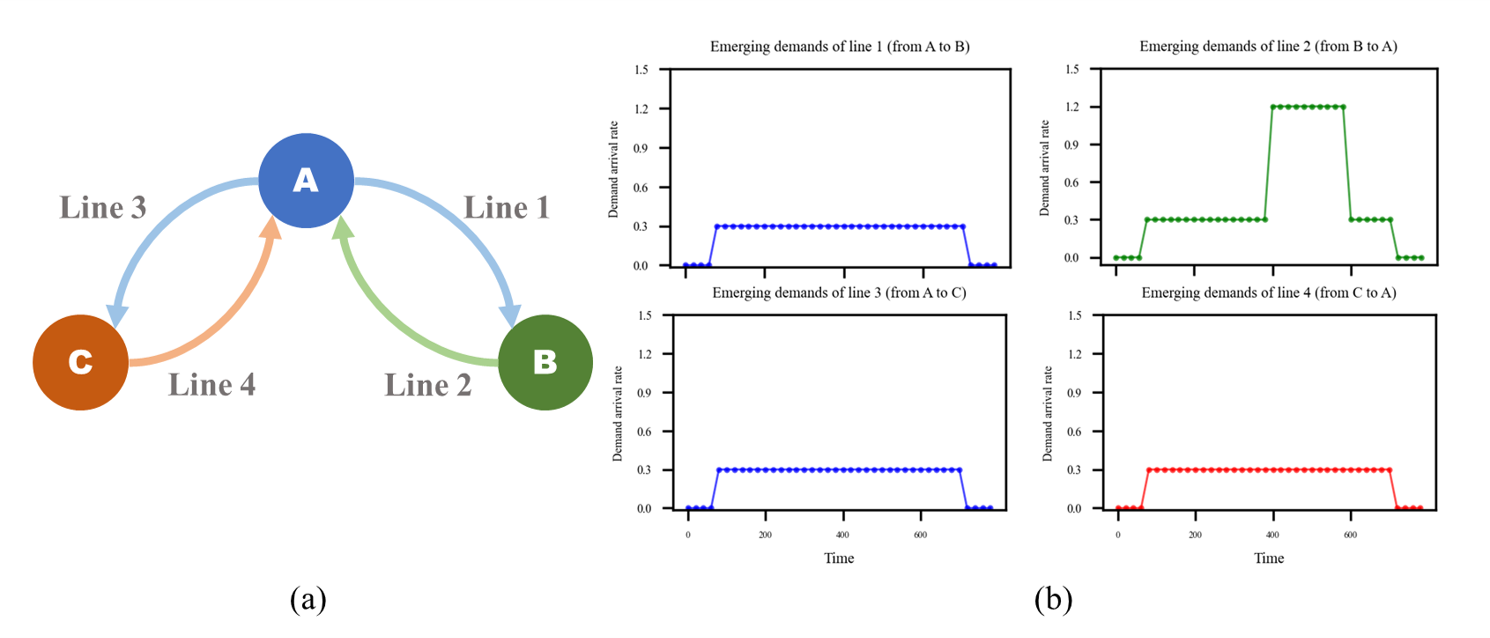}
  \caption{The 3-city-4-line toy network. (a) Topology of the network. (b) Demand of all lines.}
  \label{fig:toy2}
\end{figure}

\noindent The 3-city-4-line toy network contains one central city (A) and two neighboring cities (B, C). At the beginning of each horizon, the idle vehicles in the central city can be assigned to lines to any of the neighboring cities immediately or continue to wait for the next dispatching. Compared to experiments conducted in the 2-city-2-line network in subsection \ref{toy1}, this subsection is devoted to demonstrating the effectiveness of the proposed framework on spatially allocating vehicle resources. \textcolor{}{Figure \ref{fig:toy2}(a) depicts the network topology, where the distance from city A to city B and C is both 60 km, and the trip fares are set to 30 for all lines.} Line 1 and line 3 both originate from city A, while line 2 and line 4 originate from city B and city C respectively. Figure \ref{fig:toy2}(b) depicts the demand fluctuation on four lines. Among the four lines, a demand peak is observed during the evening periods specifically from city B to city A, whereas the other three lines experience relatively low demand throughout the day. 
In this series of experiments, a fleet of 30 homogeneous vehicles is considered.

\noindent Based on the network topology shown in Figure \ref{fig:toy2}(a), we design 5 supply and demand scenarios as follows to evaluate the spatial resource allocation performance.
\textcolor{}{
\begin{itemize}
    \item [1. ] All vehicles are initially located in city A and the demand pattern is demonstrated in Figure \ref{fig:toy2}(b).
    \item [2. ] All vehicles are initially located in city A. There also exists a demand peak on line 4 based on demands demonstrated in Figure \ref{fig:toy2}(b), but the intensity is half of the demand peak on line 2.
    \item [3. ] All vehicles are initially located in city A. The demand fluctuation is based on demands in scenario (2) while the demand peak occurs at the same time on lines 1 and 3 instead of lines 2 and 4.
    \item [4. ] The demand pattern is the same as Figure \ref{fig:toy2}(b), while all vehicles are initially located in city B.
    \item [5. ] The demand pattern is the same as Figure \ref{fig:toy2}(b), while all vehicles are initially evenly distributed in three cities. 
\end{itemize}
}
\begin{figure}[!ht]
  \centering
  \includegraphics[width=1\linewidth]{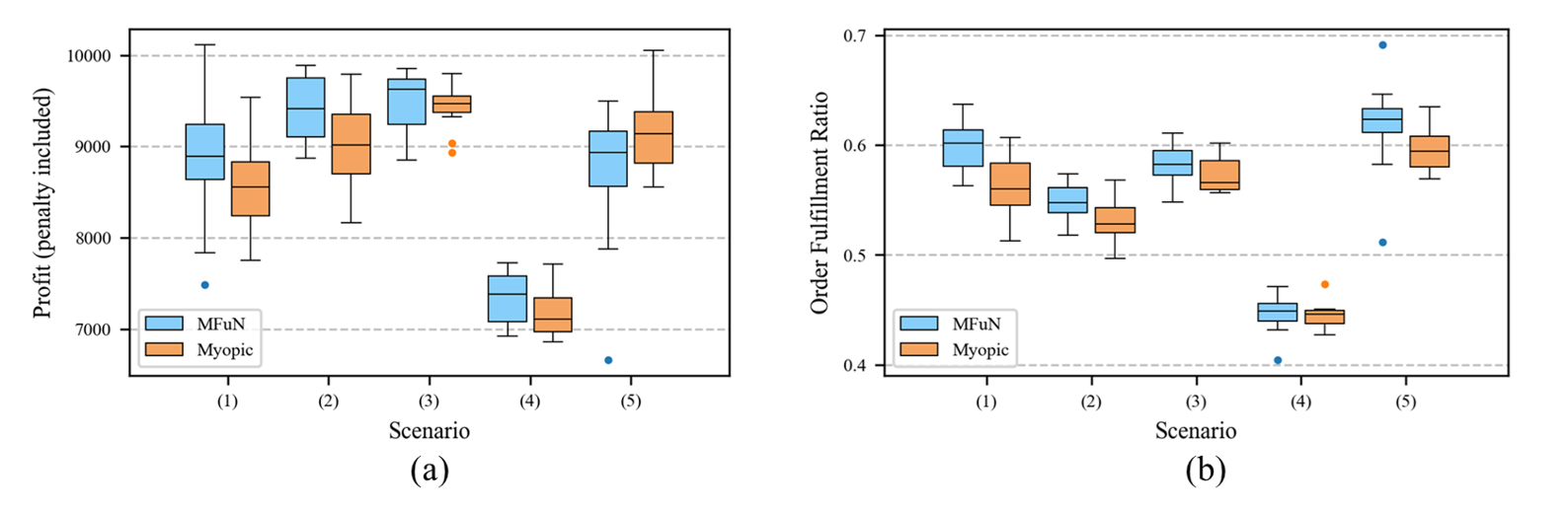}
  \caption{\textcolor{}{Performance of MFuN and Myopic in the 3-city-4-line toy network under five demand and supply scenarios. (a) Profit. (b) Order fulfillment ratio.}}
  \label{fig:d_toy2}
\end{figure}

\begin{figure}[!ht]
  \centering
  \includegraphics[width=1\linewidth]{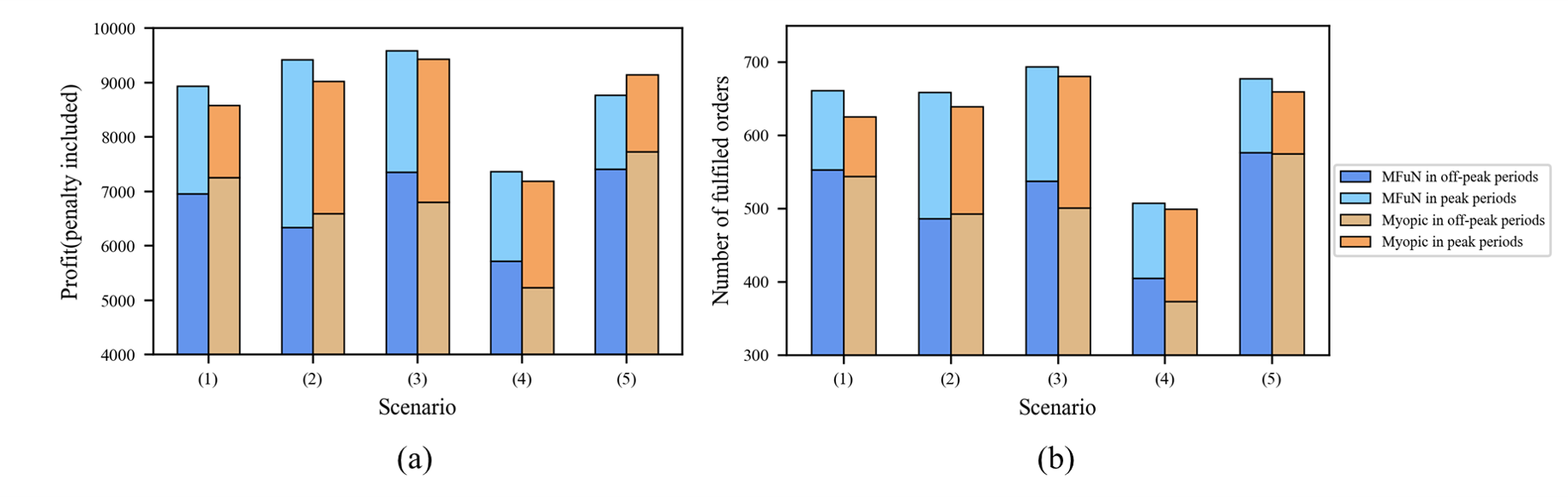}
  \caption{\textcolor{}{Performance of MFuN and Myopic in the 3-city-4-line toy network during peak and off-peak periods under five demand and supply scenarios. (a) Profit. (b)Number of fulfilled orders.}}
  \label{fig:de1_toy2}
\end{figure}

\noindent \textcolor{}{Figures \ref{fig:d_toy2} are box plots on daily profits and order fulfillment ratio of MFuN and Myopic method under five scenarios. The daily revenue of MFuN is relatively higher than the values of Myopic among the most supply and demand scenarios, basically attaining 3$\%$ to 7$\%$ improvement compared to the Myopic method. MFuN also maintains a relatively satisfactory order fulfillment ratio compared to the Myopic method.}

\noindent However, the optimization mechanism differs across the five scenarios. \textcolor{}{Figures \ref{fig:de1_toy2} illustrate the average daily profit and the average number of fulfilled orders of two methods in peak periods and off-peak periods under various scenarios.} 
Under scenarios (1) and (2), the advantage of MFuN is mainly reflected in serving more orders during high-demand periods. The MFuN framework dispatches more vehicles to city B compared to city C before and during the peak periods, which is the origination of the high-demand line. Before and during peak periods, the MFuN framework dispatches a larger number of vehicles to city B, which is the origin of the high-demand line. This allocation strategy of vehicle resources has a slight impact on service capacity during off-peak periods but ensures a relatively abundant availability of vehicles during the demand peak. There are 17$\%$ more vehicle departures from city B to fulfill 20$\%$ more orders in high-demand lines during the peak periods under the dispatching of the MFuN framework compared to the Myopic method.

\noindent In scenarios (3) and (4), the advantage of the MFuN framework during peak periods is less apparent. However, MFuN excels in providing more services primarily in low-demand lines during off-peak periods, leading to improved daily profits. This is primarily attributed to the demand peak under scenarios (3) and (4) occurring on lines originating from cities where vehicles are initially concentrated.
Under the dispatching strategy of the Myopic method, fleet operations are insufficient in the early periods of the day. As a result, a certain number of idle vehicles accumulate at the initial city during low-demand periods before the peak demand emerges. In contrast, the MFuN framework proactively dispatches more vehicles to the neighboring cities in the early stages, which exhibits more advantages in terms of fulfilling orders in low-demand lines during the off-peak periods. 

\noindent Additionally, when vehicles are evenly distributed initially under scenario (5), no significant advantages are observed for the performance of the MFuN strategies compared to the Myopic method. In this particular scenario, MFuN incurs higher traveling costs, resulting in slightly lower overall profits compared to the Myopic method.

\noindent This series of experiments highlights that the optimization effect of the MFuN framework lies primarily in its ability to exploit flexible vehicle resource allocation to mitigate spatial imbalances between supply and demand. The MFuN framework excels in scenarios where there is a significant imbalance between the demand on a specific line and the supply in its originating city, such as during peak periods in scenarios (1) and (2). In these cases, MFuN demonstrates clear advantages in efficiently dispatching fleets to mitigate the effects of supply shortages. However, in other scenarios, the advantages of the MFuN framework become less apparent due to the relatively balanced supply and demand conditions. These results highlight the adaptability of the MFuN framework to address spatial imbalances and effectively allocate resources where they are most needed. By strategically dispatching vehicles and optimizing fleet operations, MFuN can mitigate the negative impacts of supply shortages and enhance the overall performance of intercity ride-pooling services.

\subsection{Realistic network} \label{real}
\begin{figure}[!ht]
  \centering
  \includegraphics[width=0.5\linewidth]{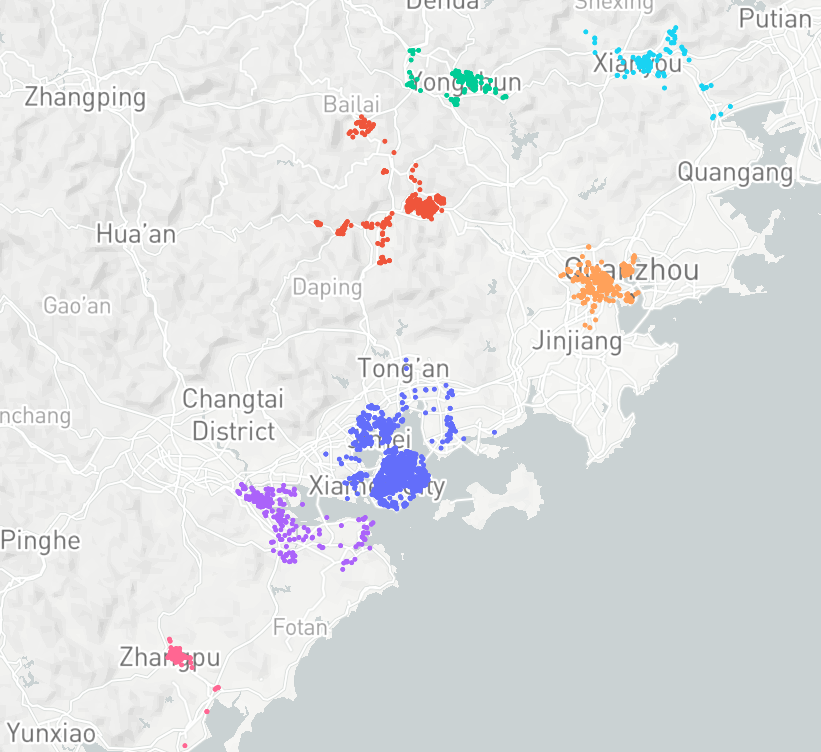}
  \caption{Distribution of realistic demand origins and destinations within an hour.}
  \label{fig:real}
\end{figure}

\noindent After verifying the effectiveness of the proposed framework from the perspective of temporally vehicle scheduling and spatial fleet allocation in subsection \ref{toy1} and \ref{toy2} respectively, we conduct numerical experiments based on a realistic network of Xiamen and its surrounding cities in this subsection.
The Xiamen city cluster is situated on the southeast coast of China. The core urban area, Xiamen Island, has a developed economy and thriving market, which is closely interconnected with surrounding cities, forming a network of transportation and economic activities. For our study, we obtain the operational data from an intercity ride-pooling platform operating in Xiamen during June and July 2022. The dataset encompasses information from 130,112 orders across 12 different lines connecting Xiamen with six surrounding cities. The surrounding cities vary in the distance to Xiamen Island. The city farthest from Xiamen is Xianyou, which is 169 kilometers away, while the nearest city Longhai is only 37 kilometers away. To cater to the intercity travel demands within the city cluster, a fleet of 291 vehicles is deployed and operated for intercity ride-pooling services in reality.
Figure \ref{fig:real} shows the distribution of origins and destinations of all intercity ride-pooling orders within an hour in the operational data.

\noindent In this subsection, we replicate the realistic operational settings by utilizing a simulator, and we also generate the training dataset based on the passenger behavior observed in the realistic demand data by employing several data-driven techniques. First, we calculate the average intensity of newly emerging demands for each line within every 10 minutes, based on the realistic demand data. \textcolor{}{Using these time-dependent parameters, we generate orders in the simulation environment by modeling the order arrival process as a non-homogeneous Poisson process. Additionally, we project the latitude and longitude of each realistic order's origin and destination into Cartesian coordinates, from which we sample the location information for orders in the simulator}. The travel time between origin and destination is calculated using the Euclidean distance metric. However, to account for the detouring behavior observed in the operational data, we calibrate the detour ratio in urban areas and on intercity roads respectively based on realistic trajectory data. This adjustment ensures that the simulated travel times reflect the real-world detouring patterns. We also generate a hard time window for picking each order. The width of the pick-up time window follows a normal distribution with a mean of 60 minutes. To ensure a satisfactory passenger experience and avoid excessive detours, we set a \textcolor{}{latest} arrival time for each order.

\noindent \textcolor{}{The one-way pooled-ride trip fare of each line in the simulator is consistent with the average level in reality, ranging from 35 to 100 for different lines. The fleet size in the simulation environment is also consistent with the average level in reality.
Due to significant variations in intercity travel demand on different days of the week, drivers' attendance rates and times to start work also exhibit differences across days. We set the number of vehicles starting service from different cities during each period of the day based on the average situation from real data. In the subsequent experiments, we used a fleet size of 192 vehicles for Monday experiments, 176 vehicles for Wednesday experiments, 215 vehicles for Friday experiments, and 221 vehicles for Sunday experiments.
By utilizing the ALNS hyper-parameters settings during training, the average total computational time for each episode is 34.1 seconds. Notice that in the experiments of realistic scenarios, each episode consists of 60 horizons of vehicle dispatching and routing for 12 lines, so on average 0.047 seconds suffice to complete one iteration of fleet dispatching and vehicle routing for a single line. This indicates that from the perspective of computational time, our algorithm has ample potential to be applied in real-time operations.
}

\subsubsection{Performance under different demand and supply scenarios}\label{de_real}

\begin{figure}[!ht]
  \centering
  \includegraphics[width=0.8\linewidth]{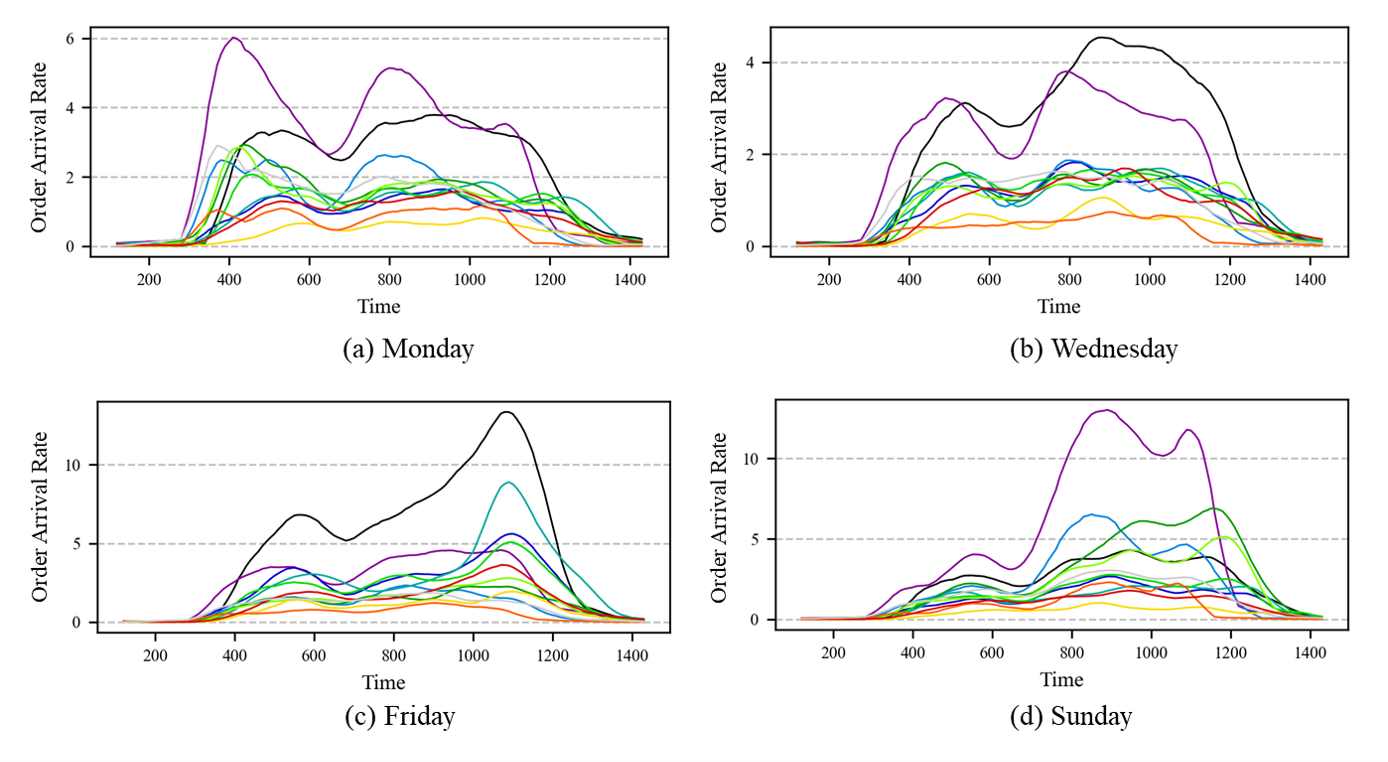}
  \caption{Daily fluctuation of demand for each line. (a) Monday. (b) Wednesday. (c) Friday. (d) Sunday.}
  \label{fig:demand_flu_real}
\end{figure}

\begin{figure}[!ht]
  \centering
  \includegraphics[width=1\linewidth]{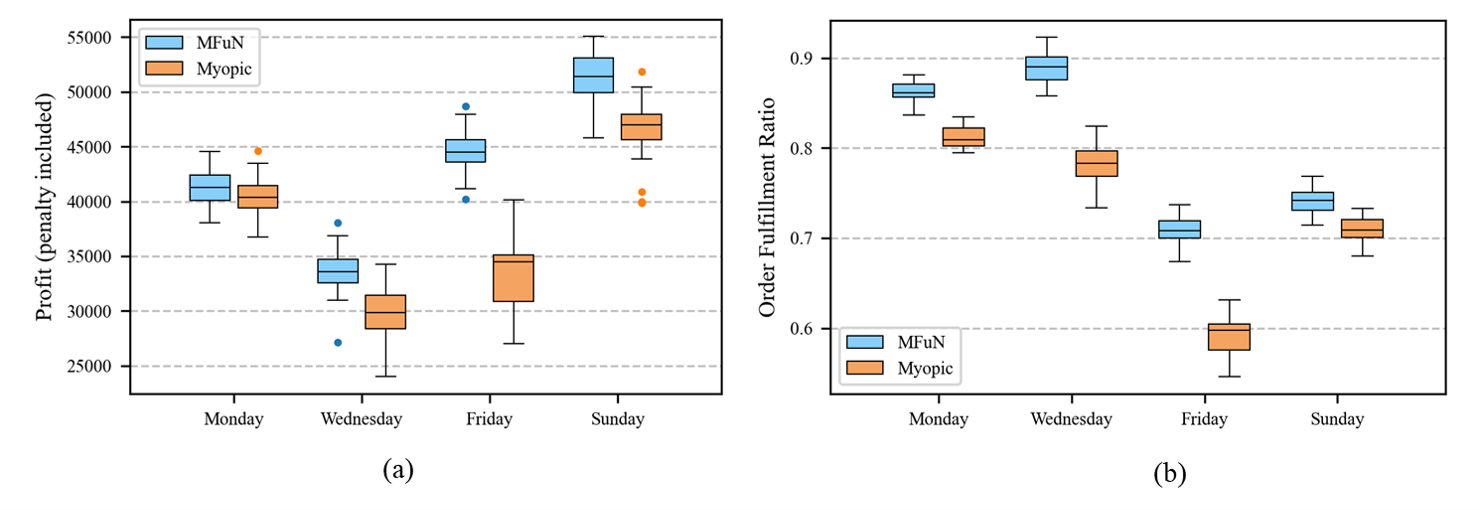}
  \caption{Performance of MFuN and Myopic in the realistic network. (a) Profit. (b) Order fulfillment ratio.}
  \label{fig:de_real}
\end{figure}

\noindent \textcolor{}{The operational data reveals that the intercity travel demands for each line follow a weekly cycle. The travel demands on the same weekday of each week exhibit similar patterns, while those on different weekdays show significant variations}. Figure \ref{fig:demand_flu_real} illustrates the daily fluctuations of emerging order numbers per 10 minutes for 12 lines on  Monday, Wednesday, Friday, and Sunday. For each week, there is a substantial demand from surrounding cities to the central city on Sunday afternoons and early Monday mornings. Conversely, on Friday afternoons, there is an almost equal amount of demand from the central city to surrounding cities. There also exists demand peaks in the morning and evening of each day, although these peaks have lower intensity compared to the weekly fluctuation pattern. Furthermore, it's worth noting that the demand varies in scale across different lines, as shown by different \textcolor{}{curves} in Figure \ref{fig:demand_flu_real}. The complex demand patterns contribute to significant spatiotemporal variations in both demand and supply. The ride-pooling platform must intelligently allocate vehicle resources among the lines and optimize the matching of supply and demand through flexible scheduling, taking into account the specific temporal characteristics of the demand patterns. We train models for fleet operations under the demand scenarios of four different weekdays respectively to illustrate the effectiveness of the proposed MFuN framework, as shown in Figure \ref{fig:de_real}.

\noindent It is observed that the MFuN outperforms the Myopic method under demand and supply patterns of various weekdays, attaining improvement on both average daily profit and order fulfillment ratio. The advantages of MFuN are particularly pronounced when there is a significant imbalance between demand and supply, such as during peak periods \textcolor{}{on} Fridays. In this scenario, MFuN demonstrates a remarkable 27$\%$ improvement in daily profits compared to the Myopic method. Under the scenario of Monday where the demand peak emerges at the beginning of a day, MFuN may not have enough time to effectively allocate resources to accommodate the sudden surge in demand. Under the Wednesday scenario, MFuN attains satisfactory order fulfillment rates when demand is relatively sparse compared to the weekend.

\begin{figure}[!ht]
  \centering
  \includegraphics[width=0.8\linewidth]{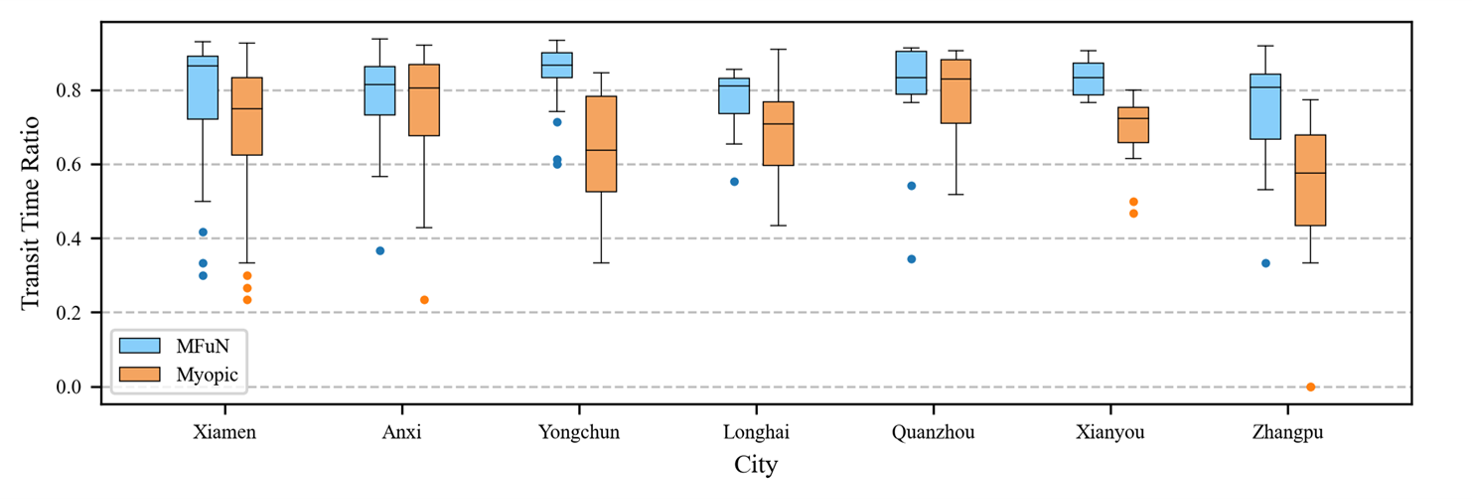}
  \caption{Average percentage of time in transit for vehicles originating from each city under the Sunday scenario.}
  \label{fig:time_transit_real}
\end{figure}

\begin{figure}[!ht]
  \centering
  \includegraphics[width=0.8\linewidth]
  {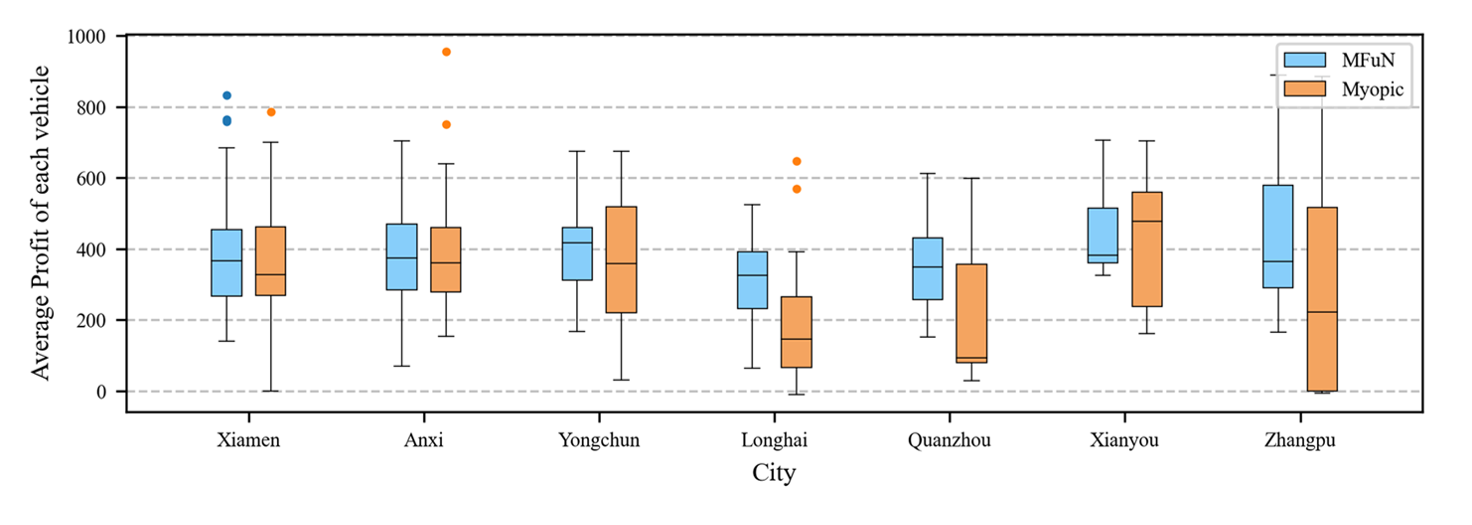}
  \caption{Average daily revenue for drivers originating from each city under the Sunday scenario.}
  \label{fig:average_revenue_real}
\end{figure}

\noindent The advantage of MFuN is reflected in the intelligent spatiotemporal allocation of vehicle resources. From a spatial perspective, MFuN dispatches all available resources in the system in a more flexible manner. Figure \ref{fig:time_transit_real} shows the percentage of time in transit for vehicles originating from each city under the Sunday scenario. Under the MFuN dispatching strategy, vehicles from different cities have similar transit times, indicating a balanced allocation of resources. On the other hand, under the Myopic dispatching approach, vehicles originating from cities with lower demand, such as Yongchun and Zhangpu, may experience longer idle times. 
The proactive dispatching strategy employed by MFuN leads to improved overall utilization of vehicle resources, which aligns with the platform operator's expectations. Furthermore, the revenue distribution among drivers is also more balanced under the MFuN framework, as demonstrated in Figure \ref{fig:average_revenue_real}, which indicates a more systematic task assignment facilitated by MFuN. These improvements highlight the effectiveness of agent cooperation within the multi-agent feudal networks, further enhancing the operational outcomes of the ride-pooling platform.

\begin{figure}[!ht]
  \centering
  \includegraphics[width=0.8\linewidth]{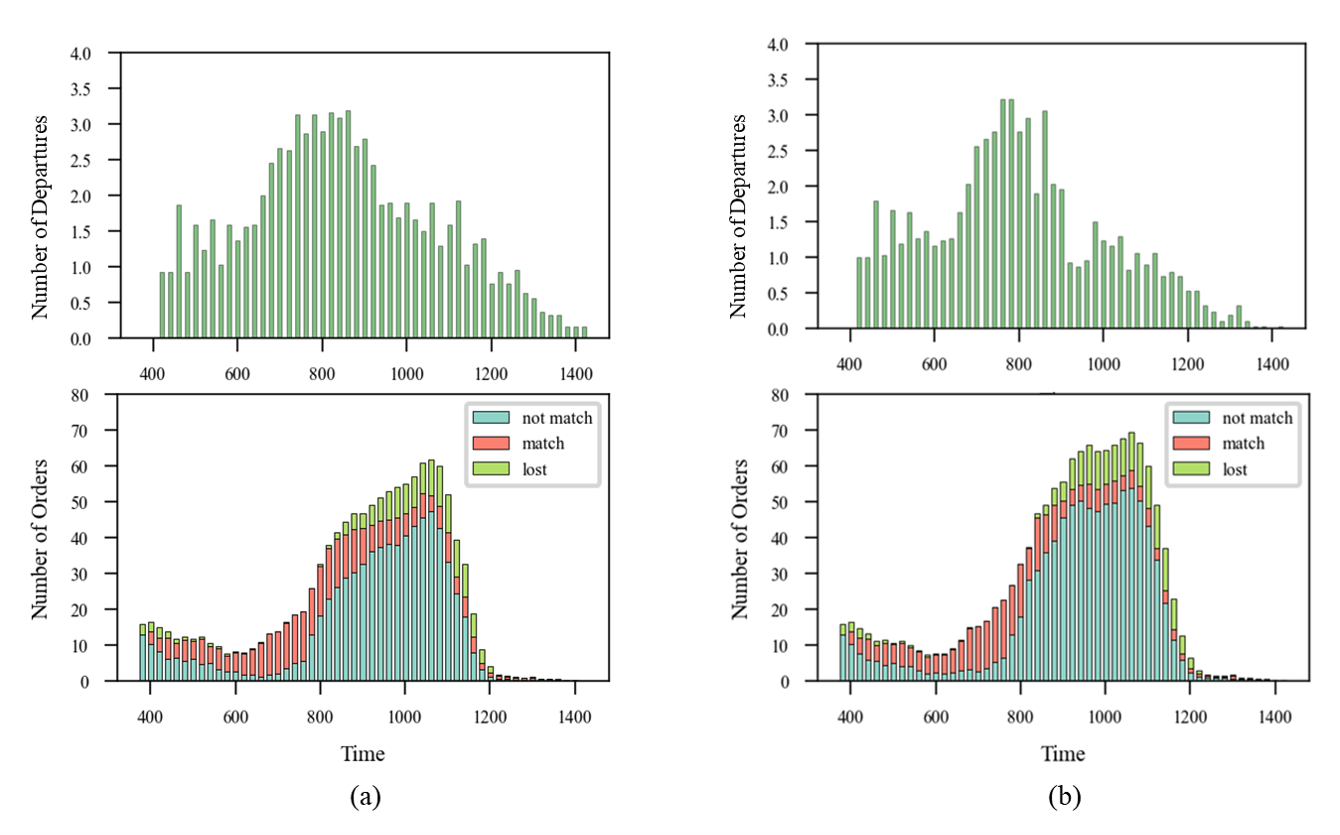}
  \caption{Number of vehicle departures and orders for a line in each period. (a) MFuN. (b) Myopic.}
  \label{fig:ex_real}
\end{figure}

\noindent From a temporal scheduling perspective, MFuN demonstrates its ability to identify and respond to potential fluctuations in demand and supply within specific cities (agents). This adaptability allows MFuN to make informed decisions based on anticipated dynamics. Figure \ref{fig:ex_real} presents the number of departures and orders for a line with a demand peak in the afternoon.
Under the MFuN framework, the demand peak of this line is recognized, prompting the dispatching of a greater number of vehicles to the originating city. As a result, vehicles can depart continuously during peak periods to fulfill orders. In contrast, the Myopic method dispatches only a limited number of vehicles to the originating city during the peak periods, leading to a shortage of available vehicles and a higher rate of order loss in subsequent periods.

\subsubsection{Sensitivity analysis of supply and demand level} \label{s_real}
\begin{figure}[!ht]
  \centering
  \includegraphics[width=1\linewidth]{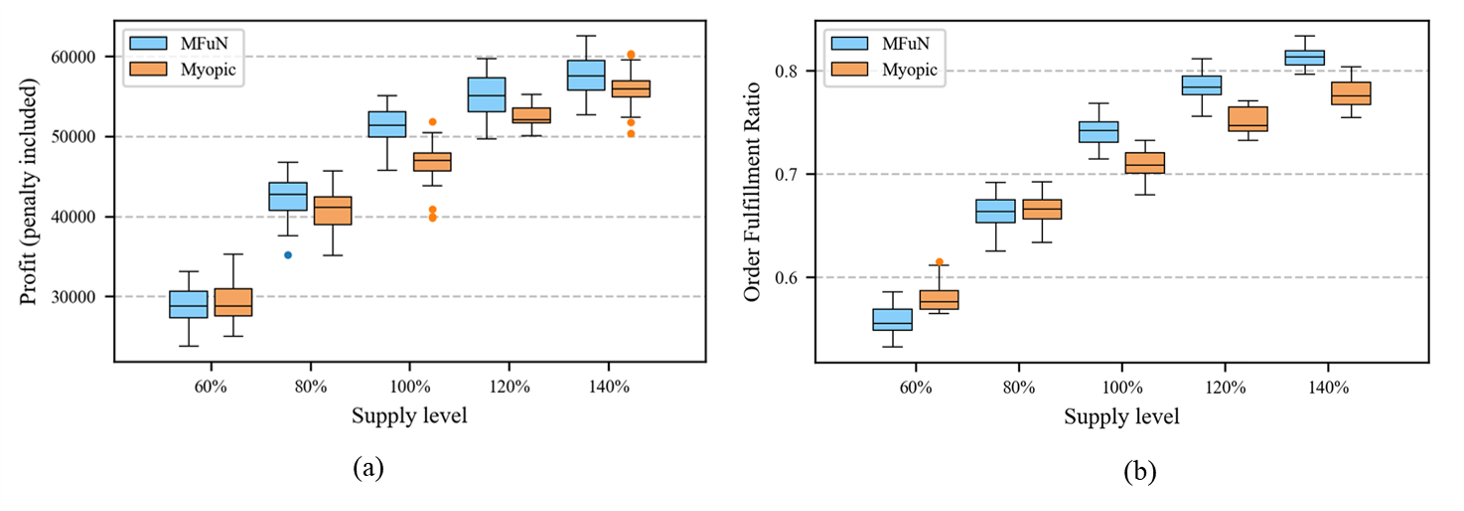}
  \caption{Performance of MFuN and Myopic under different supply levels. (a) Profit. (b) Order fulfillment ratio.}
  \label{fig:s_real}
\end{figure}

\begin{figure}[!ht]
  \centering
  \includegraphics[width=1\linewidth]{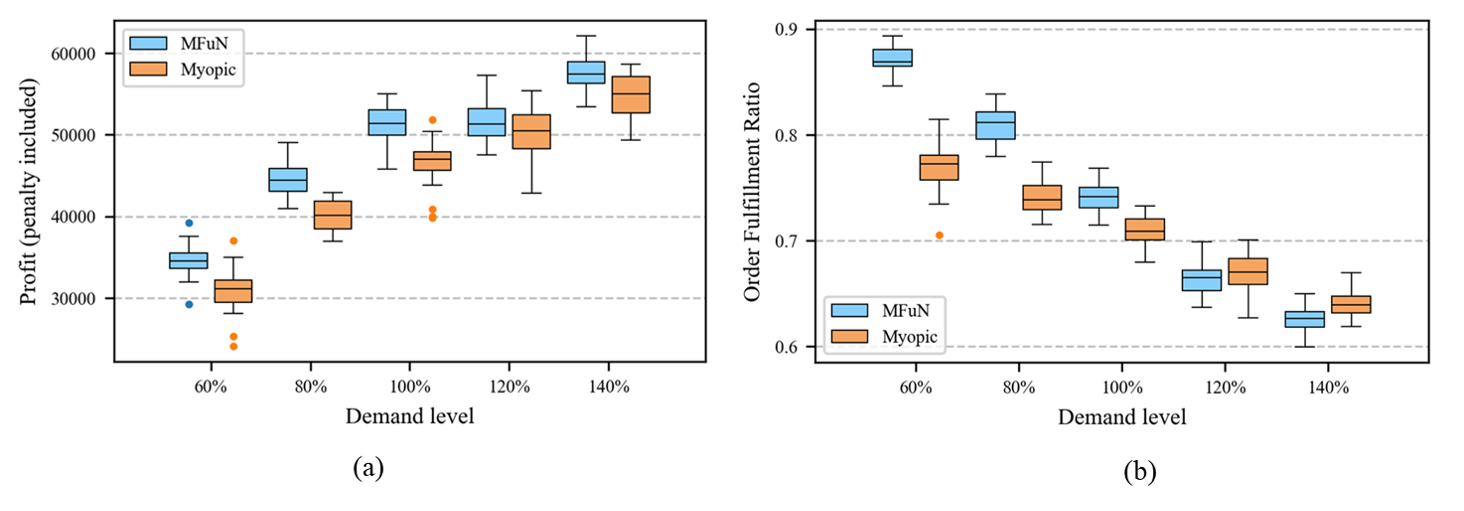}
  \caption{Performance of MFuN and Myopic under different demand levels. (a) Profit. (b) Order fulfillment ratio.}
  \label{fig:d_real}
\end{figure}

\noindent \textcolor{}{In this subsection, we perform sensitivity analyses to evaluate the performance of fleet operational strategies under varying supply and demand levels}.
The experiments on different supply levels vary on the number of vehicles applied in each line. Five supply quantity levels are set, ranging from 60$\%$ to 140$\%$ based on the fleet under the Sunday scenario in realistic operational data. The performance of both methods in terms of average daily profit and order fulfillment ratio is presented in Figure \ref{fig:s_real}. A series of experiments are also conducted to explore the impact of different demand densities on the performance of the fleet operational strategies. The demand densities vary from shrinking by 40$\%$ to amplifying by 40$\%$ based on the demand dynamics under the Sunday scenario in realistic operational data. The performance comparisons under various demand levels are shown in Figure \ref{fig:d_real}.

\noindent These two series of experiments reveal consistent patterns in the performance of the two methods. The MFuN framework demonstrates superior performance over the Myopic method when the demand levels are relatively low. This advantage arises from the ability of MFuN to intelligently allocate vehicles to high-demand lines, thereby maximizing the system's profit. Even in scenarios where the overall demand level is low or the supply level is high, there are still peak periods and lines that lack sufficient vehicles. In such cases, MFuN excels in optimizing the allocation of vehicles to high-demand lines, whereas the Myopic method often results in a large number of idle vehicles in low-demand regions.

\noindent However, under conditions where the supply level is relatively low or the demand level is relatively high, vehicle resources can always be in high utilization. In these situations, the Myopic method already guarantees that vehicles are utilized to their fullest capacity, leaving little room for MFuN to further improve the matching of demand and vehicle resources.

\section{Conclusion} \label{sec_conclusion}

\noindent This study addresses the challenging problem of online fleet operations in on-demand intercity ride-pooling services within city clusters, characterized by a complex embedded structure. The fleet operations involve assigning idle vehicles in each city to lines and optimizing the vehicle routing for each line dynamically. The assignment of vehicles to intercity lines can be regarded as a stochastic dynamic vehicle resource allocation problem, while the online order-matching and vehicle routing process is a variant of dynamic DARP. The integrated online optimization framework of vehicle resource allocation and dynamic vehicle routing is barely touched in the literature. 

\noindent To address this gap, we propose a two-level framework that combines reinforcement learning and heuristic approaches. A novel multi-agent hierarchical reinforcement learning method named multi-agent feudal networks is adopted in the upper level of the framework. Each worker agent corresponds to a city and collaboratively makes decisions on vehicle assignments under the guidance of a manager agent. The outputs of the worker agents, represented in a multi-discrete manner, are interpreted and mapped by a mixed-integer linear programming (MILP) formulation that considers the heterogeneity of vehicles. In the lower level of the framework, we solve the dynamic DARP using an adaptive large neighborhood search heuristic based on the assigned vehicles for each line.

\noindent We conduct extensive numerical experiments to evaluate the effectiveness of our proposed framework based on two toy networks and a realistic network in Xiamen. The experiments on the toy networks demonstrate the efficacy of our framework in temporal scheduling and spatial allocation. Furthermore, our proposed framework outperforms the myopic method across various demand and supply scenarios in the realistic network with 7 cities and 12 lines, attaining improvements in average daily profit and order fulfillment ratio. The optimization effect of MFuN is primarily observed in mitigating the spatial-temporal imbalance between supply and demand, with relatively balanced scenarios showing fewer advantages for MFuN.

\noindent In future studies, we plan to extend the scope of our research from on-demand orders to include orders with different reservation lead times. The platform can design price mechanisms to influence the stochastic demand and improve the fleet operation strategy to provide services to hybrid demand including on-demand orders and orders reserved in advance. Another area of investigation is the planning of intercity ride-pooling services within city clusters, which encompasses determining the optimal fleet size, pricing mechanisms, and operation lines. This planning can be constructed on the market equilibrium of intercity passenger transport, taking into account the competition with other intercity transit modes.

\bibliographystyle{apalike}
\bibliography{ref}

\begin{thebibliography}{}

\bibitem[Agatz et~al., 2012]{agatz2012optimization}
Agatz, N., Erera, A., Savelsbergh, M., and Wang, X. (2012).
\newblock Optimization for dynamic ride-sharing: A review.
\newblock {\em European Journal of Operational Research}, 223(2):295--303.

\bibitem[Ahilan and Dayan, 2019]{ahilan2019feudal}
Ahilan, S. and Dayan, P. (2019).
\newblock Feudal multi-agent hierarchies for cooperative reinforcement learning.
\newblock {\em arXiv preprint arXiv:1901.08492}.

\bibitem[{Alonso-Mora} et~al., 2017]{alonso-moraPredictiveRoutingAutonomous2017}
{Alonso-Mora}, J., Wallar, A., and Rus, D. (2017).
\newblock Predictive routing for autonomous mobility-on-demand systems with ride-sharing.
\newblock In {\em 2017 {{IEEE}}/{{RSJ International Conference}} on {{Intelligent Robots}} and {{Systems}} ({{IROS}})}, pages 3583--3590, {Vancouver, BC}. {IEEE}.

\bibitem[Bongiovanni et~al., 2022]{bongiovanni2022machine}
Bongiovanni, C., Kaspi, M., Cordeau, J.-F., and Geroliminis, N. (2022).
\newblock A machine learning-driven two-phase metaheuristic for autonomous ridesharing operations.
\newblock {\em Transportation Research Part E: Logistics and Transportation Review}, 165:102835.

\bibitem[Braekers and Kovacs, 2016]{braekers2016multi}
Braekers, K. and Kovacs, A.~A. (2016).
\newblock A multi-period dial-a-ride problem with driver consistency.
\newblock {\em Transportation Research Part B: Methodological}, 94:355--377.

\bibitem[Cordeau, 2006]{Cordeau2006}
Cordeau, J. (2006).
\newblock A branch-and-cut algorithm for the dial-a-ride problem.
\newblock {\em Operations Research}, 54(4):573--586.

\bibitem[Czioska et~al., 2019]{czioska2019real}
Czioska, P., Kutadinata, R., Trifunovi{\'c}, A., Winter, S., Sester, M., and Friedrich, B. (2019).
\newblock Real-world meeting points for shared demand-responsive transportation systems.
\newblock {\em Public Transport}, 11:341--377.

\bibitem[Dayan and Hinton, 1992]{dayan1992feudal}
Dayan, P. and Hinton, G.~E. (1992).
\newblock Feudal reinforcement learning.
\newblock {\em Advances in neural information processing systems}, 5.

\bibitem[Demir et~al., 2012]{Demir2012}
Demir, E., Bektaş, T., and Laporte, G. (2012).
\newblock An adaptive large neighborhood search heuristic for the pollution-routing problem.
\newblock {\em European Journal of Operational Research}, 223(2):346--359.

\bibitem[Fang and Yu, 2017]{fang2017urban}
Fang, C. and Yu, D. (2017).
\newblock Urban agglomeration: An evolving concept of an emerging phenomenon.
\newblock {\em Landscape and urban planning}, 162:126--136.

\bibitem[Fielbaum et~al., 2021]{fielbaum2021demand}
Fielbaum, A., Bai, X., and Alonso-Mora, J. (2021).
\newblock On-demand ridesharing with optimized pick-up and drop-off walking locations.
\newblock {\em Transportation research part C: emerging technologies}, 126:103061.

\bibitem[Ganji et~al., 2021]{GANJI2021345}
Ganji, S., Ahangar, A., Awasthi, A., and {Jamshidi Bandari}, S. (2021).
\newblock Psychological analysis of intercity bus passenger satisfaction using q methodology.
\newblock {\em Transportation Research Part A: Policy and Practice}, 154:345--363.

\bibitem[Ghilas et~al., 2016]{Ghilas2016}
Ghilas, V., Demir, E., and Van~Woensel, T. (2016).
\newblock An adaptive large neighborhood search heuristic for the pickup and delivery problem with time windows and scheduled lines.
\newblock {\em Computers \& Operations Research}, 72:12--30.

\bibitem[Godfrey and Powell, 2002]{godfrey2002adaptive}
Godfrey, G.~A. and Powell, W.~B. (2002).
\newblock An adaptive dynamic programming algorithm for dynamic fleet management, ii: Multiperiod travel times.
\newblock {\em Transportation Science}, 36(1):40--54.

\bibitem[Goeke, 2019]{goeke2019granular}
Goeke, D. (2019).
\newblock Granular tabu search for the pickup and delivery problem with time windows and electric vehicles.
\newblock {\em European Journal of Operational Research}, 278(3):821--836.

\bibitem[Gschwind and Drexl, 2019]{Gschwind2019}
Gschwind, T. and Drexl, M. (2019).
\newblock Adaptive large neighborhood search with a constant-time feasibility test for the dial-a-ride problem.
\newblock {\em Transportation Science}, 53(2):480--491.

\bibitem[Gschwind and Irnich, 2015]{gschwind2015effective}
Gschwind, T. and Irnich, S. (2015).
\newblock Effective handling of dynamic time windows and its application to solving the dial-a-ride problem.
\newblock {\em Transportation Science}, 49(2):335--354.

\bibitem[Guo and Xu, 2020]{guo2020deep}
Guo, G. and Xu, Y. (2020).
\newblock A deep reinforcement learning approach to ride-sharing vehicle dispatching in autonomous mobility-on-demand systems.
\newblock {\em IEEE Intelligent Transportation Systems Magazine}, 14(1):128--140.

\bibitem[Guo and Zhang, 2020]{guo2020residual}
Guo, G. and Zhang, T. (2020).
\newblock A residual spatio-temporal architecture for travel demand forecasting.
\newblock {\em Transportation Research Part C: Emerging Technologies}, 115:102639.

\bibitem[Guo et~al., 2022]{guo2022vehicle}
Guo, J., Long, J., Xu, X., Yu, M., and Yuan, K. (2022).
\newblock The vehicle routing problem of intercity ride-sharing between two cities.
\newblock {\em Transportation Research Part B: Methodological}, 158:113--139.

\bibitem[Guo et~al., 2021]{guo2021robust}
Guo, X., Caros, N.~S., and Zhao, J. (2021).
\newblock Robust matching-integrated vehicle rebalancing in ride-hailing system with uncertain demand.
\newblock {\em Transportation Research Part B: Methodological}, 150:161--189.

\bibitem[Guo et~al., 2020]{gyh2020}
Guo, Y., Zhang, Y., Yu, J., and Shen, X. (2020).
\newblock A spatiotemporal thermo guidance based real-time online ride-hailing dispatch framework.
\newblock {\em IEEE Access}, 8:115063--115077.

\bibitem[Ho et~al., 2018]{Ho2018}
Ho, S.~C., Szeto, W.~Y., Kuo, Y.~H., Leung, J. M.~Y., Petering, M., and Tou, T. W.~H. (2018).
\newblock A survey of dial-a-ride problems: Literature review and recent developments.
\newblock {\em Transportation Research Part B-Methodological}, 111:395--421.

\bibitem[Huang et~al., 2020]{huang2020two}
Huang, D., Gu, Y., Wang, S., Liu, Z., and Zhang, W. (2020).
\newblock A two-phase optimization model for the demand-responsive customized bus network design.
\newblock {\em Transportation Research Part C: Emerging Technologies}, 111:1--21.

\bibitem[Jiao et~al., 2021]{jiao2021real}
Jiao, Y., Tang, X., Qin, Z.~T., Li, S., Zhang, F., Zhu, H., and Ye, J. (2021).
\newblock Real-world ride-hailing vehicle repositioning using deep reinforcement learning.
\newblock {\em Transportation Research Part C: Emerging Technologies}, 130:103289.

\bibitem[Jin et~al., 2019]{jin2019coride}
Jin, J., Zhou, M., Zhang, W., Li, M., Guo, Z., Qin, Z., Jiao, Y., Tang, X., Wang, C., Wang, J., et~al. (2019).
\newblock Coride: joint order dispatching and fleet management for multi-scale ride-hailing platforms.
\newblock In {\em Proceedings of the 28th ACM International Conference on Information and Knowledge Management}, pages 1983--1992.

\bibitem[Ke et~al., 2017]{ke2017short}
Ke, J., Zheng, H., Yang, H., and Chen, X.~M. (2017).
\newblock Short-term forecasting of passenger demand under on-demand ride services: A spatio-temporal deep learning approach.
\newblock {\em Transportation research part C: Emerging technologies}, 85:591--608.

\bibitem[Kullman et~al., 2022]{kullman2022dynamic}
Kullman, N.~D., Cousineau, M., Goodson, J.~C., and Mendoza, J.~E. (2022).
\newblock Dynamic ride-hailing with electric vehicles.
\newblock {\em Transportation Science}, 56(3):775--794.

\bibitem[Lee et~al., 2021]{lee2021zonal}
Lee, E., Cen, X., and Lo, H.~K. (2021).
\newblock Zonal-based flexible bus service under elastic stochastic demand.
\newblock {\em Transportation Research Part E: Logistics and Transportation Review}, 152:102367.

\bibitem[Lei et~al., 2020]{Lei2020}
Lei, C., Jiang, Z.~T., and Ouyang, Y.~F. (2020).
\newblock Path-based dynamic pricing for vehicle allocation in ridesharing systems with fully compliant drivers.
\newblock {\em Transportation Research Part B-Methodological}, 132:60--75.

\bibitem[Liu and Ouyang, 2021]{liu2021mobility}
Liu, Y. and Ouyang, Y. (2021).
\newblock Mobility service design via joint optimization of transit networks and demand-responsive services.
\newblock {\em Transportation Research Part B: Methodological}, 151:22--41.

\bibitem[Liu et~al., 2022]{liu2022deep}
Liu, Y., Wu, F., Lyu, C., Li, S., Ye, J., and Qu, X. (2022).
\newblock Deep dispatching: A deep reinforcement learning approach for vehicle dispatching on online ride-hailing platform.
\newblock {\em Transportation Research Part E: Logistics and Transportation Review}, 161:102694.

\bibitem[Lowe et~al., 2017]{lowe2017multi}
Lowe, R., Wu, Y.~I., Tamar, A., Harb, J., Pieter~Abbeel, O., and Mordatch, I. (2017).
\newblock Multi-agent actor-critic for mixed cooperative-competitive environments.
\newblock {\em Advances in neural information processing systems}, 30.

\bibitem[Luo et~al., 2019]{luo2019two}
Luo, Z., Liu, M., and Lim, A. (2019).
\newblock A two-phase branch-and-price-and-cut for a dial-a-ride problem in patient transportation.
\newblock {\em Transportation Science}, 53(1):113--130.

\bibitem[Ma and Koutsopoulos, 2022]{maNearondemandMobilityBenefits2022a}
Ma, Z. and Koutsopoulos, H.~N. (2022).
\newblock Near-on-demand mobility. {{The}} benefits of user flexibility for ride-pooling services.
\newblock {\em Transportation Research Part C: Emerging Technologies}, 135:103530.

\bibitem[Mao et~al., 2020]{mao2020dispatch}
Mao, C., Liu, Y., and Shen, Z.-J.~M. (2020).
\newblock Dispatch of autonomous vehicles for taxi services: A deep reinforcement learning approach.
\newblock {\em Transportation Research Part C: Emerging Technologies}, 115:102626.

\bibitem[Melis and S{\"o}rensen, 2022a]{melis2022real}
Melis, L. and S{\"o}rensen, K. (2022a).
\newblock The real-time on-demand bus routing problem: The cost of dynamic requests.
\newblock {\em Computers \& Operations Research}, 147:105941.

\bibitem[Melis and S{\"o}rensen, 2022b]{melis2022static}
Melis, L. and S{\"o}rensen, K. (2022b).
\newblock The static on-demand bus routing problem: large neighborhood search for a dial-a-ride problem with bus station assignment.
\newblock {\em International Transactions in Operational Research}, 29(3):1417--1453.

\bibitem[MOT, 2022]{tr2021}
MOT (2022).
\newblock Statistical communiqué of the people's republic of china on the transport industry.
\newblock \url{https://xxgk.mot.gov.cn/2020/jigou/zhghs/202205/t20220524_3656659.html }.
\newblock Accessed May 25, 2022.

\bibitem[Naccache et~al., 2018]{naccache2018multi}
Naccache, S., C{\^o}t{\'e}, J.-F., and Coelho, L.~C. (2018).
\newblock The multi-pickup and delivery problem with time windows.
\newblock {\em European Journal of Operational Research}, 269(1):353--362.

\bibitem[Ouyang et~al., 2021]{ouyang2021performance}
Ouyang, Y., Yang, H., and Daganzo, C.~F. (2021).
\newblock Performance of reservation-based carpooling services under detour and waiting time restrictions.
\newblock {\em Transportation Research Part B: Methodological}, 150:370--385.

\bibitem[Qin et~al., 2021a]{qin2021optimizing}
Qin, G., Luo, Q., Yin, Y., Sun, J., and Ye, J. (2021a).
\newblock Optimizing matching time intervals for ride-hailing services using reinforcement learning.
\newblock {\em Transportation Research Part C: Emerging Technologies}, 129:103239.

\bibitem[Qin et~al., 2021b]{qin2021multi}
Qin, X., Yang, H., Wu, Y., and Zhu, H. (2021b).
\newblock Multi-party ride-matching problem in the ride-hailing market with bundled option services.
\newblock {\em Transportation Research Part C: Emerging Technologies}, 131:103287.

\bibitem[Qin et~al., 2020]{qin2020ride}
Qin, Z., Tang, X., Jiao, Y., Zhang, F., Xu, Z., Zhu, H., and Ye, J. (2020).
\newblock Ride-hailing order dispatching at didi via reinforcement learning.
\newblock {\em INFORMS Journal on Applied Analytics}, 50(5):272--286.

\bibitem[Ropke and Pisinger, 2006]{Ropke2006}
Ropke, S. and Pisinger, D. (2006).
\newblock An adaptive large neighborhood search heuristic for the pickup and delivery problem with time windows.
\newblock {\em Transportation Science}, 40(4):455--472.

\bibitem[Schwieterman et~al., 2021]{schwieterman2021brink}
Schwieterman, J., Antolin, B., Bell, C., et~al. (2021).
\newblock On the brink: 2021 outlook for the intercity bus industry in the united states.

\bibitem[Simonetto et~al., 2019]{simonettoRealtimeCityscaleRidesharing2019}
Simonetto, A., Monteil, J., and Gambella, C. (2019).
\newblock Real-time city-scale ridesharing via linear assignment problems.
\newblock {\em Transportation Research Part C: Emerging Technologies}, 101:208--232.

\bibitem[Sun et~al., 2020a]{SunP2020}
Sun, P., Veelenturf, L.~P., Hewitt, M., and Van~Woensel, T. (2020a).
\newblock Adaptive large neighborhood search for the time-dependent profitable pickup and delivery problem with time windows.
\newblock {\em Transportation Research Part E-Logistics and Transportation Review}, 138.

\bibitem[Sun et~al., 2020b]{sun2020optimizing}
Sun, Q., Chien, S., Hu, D., Chen, G., and Jiang, R.-S. (2020b).
\newblock Optimizing multi-terminal customized bus service with mixed fleet.
\newblock {\em IEEE Access}, 8:156456--156469.

\bibitem[Tafreshian et~al., 2021]{tafreshian2021proactive}
Tafreshian, A., Abdolmaleki, M., Masoud, N., and Wang, H. (2021).
\newblock Proactive shuttle dispatching in large-scale dynamic dial-a-ride systems.
\newblock {\em Transportation Research Part B: Methodological}, 150:227--259.

\bibitem[Tang et~al., 2019]{tang2019deep}
Tang, X., Qin, Z., Zhang, F., Wang, Z., Xu, Z., Ma, Y., Zhu, H., and Ye, J. (2019).
\newblock A deep value-network based approach for multi-driver order dispatching.
\newblock In {\em Proceedings of the 25th ACM SIGKDD international conference on knowledge discovery \& data mining}, pages 1780--1790.

\bibitem[Tang et~al., 2020]{Tang2020}
Tang, X.~D., Li, M., Lin, X., and He, F. (2020).
\newblock Online operations of automated electric taxi fleets: An advisor-student reinforcement learning framework.
\newblock {\em Transportation Research Part C-Emerging Technologies}, 121.

\bibitem[Tong et~al., 2016]{tong2016online}
Tong, Y., She, J., Ding, B., Chen, L., Wo, T., and Xu, K. (2016).
\newblock Online minimum matching in real-time spatial data: experiments and analysis.
\newblock {\em Proceedings of the VLDB Endowment}, 9(12):1053--1064.

\bibitem[Tsai, 2020]{tsai2020self}
Tsai, T.-H. (2020).
\newblock Self-evolutionary sibling models to forecast railway arrivals using reservation data.
\newblock {\em Engineering Applications of Artificial Intelligence}, 96:103960.

\bibitem[Tuncel et~al., 2023]{tuncelIntegratedRidematchingVehiclerebalancing2023}
Tuncel, K., Koutsopoulos, H.~N., and Ma, Z. (2023).
\newblock An integrated ride-matching and vehicle-rebalancing model for shared mobility on-demand services.
\newblock {\em Computers \& Operations Research}, 159:106317.

\bibitem[Vansteenwegen et~al., 2022]{vansteenwegen_survey_2022}
Vansteenwegen, P., Melis, L., Aktaş, D., Montenegro, B. D.~G., Vieira, F.~S., and Sörensen, K. (2022).
\newblock A survey on demand-responsive public bus systems.
\newblock {\em Transportation Research Part C: Emerging Technologies}, 137:103573.

\bibitem[Vezhnevets et~al., 2017]{vezhnevets2017feudal}
Vezhnevets, A.~S., Osindero, S., Schaul, T., Heess, N., Jaderberg, M., Silver, D., and Kavukcuoglu, K. (2017).
\newblock Feudal networks for hierarchical reinforcement learning.
\newblock In {\em International Conference on Machine Learning}, pages 3540--3549. PMLR.

\bibitem[Wang and Yang, 2019]{wang2019ridesourcing}
Wang, H. and Yang, H. (2019).
\newblock Ridesourcing systems: A framework and review.
\newblock {\em Transportation Research Part B: Methodological}, 129:122--155.

\bibitem[Wu et~al., 2022]{wu2022time}
Wu, Y., Poon, M., Yuan, Z., and Xiao, Q. (2022).
\newblock Time-dependent customized bus routing problem of large transport terminals considering the impact of late passengers.
\newblock {\em Transportation Research Part C: Emerging Technologies}, 143:103859.

\bibitem[Xu et~al., 2018a]{xu2018large}
Xu, Z., Li, Z., Guan, Q., Zhang, D., Li, Q., Nan, J., Liu, C., Bian, W., and Ye, J. (2018a).
\newblock Large-scale order dispatch in on-demand ride-hailing platforms: A learning and planning approach.
\newblock In {\em Proceedings of the 24th ACM SIGKDD International Conference on Knowledge Discovery \& Data Mining}, pages 905--913.

\bibitem[Xu et~al., 2018b]{Xu2018}
Xu, Z., Li, Z.~X., Guan, Q.~W., Zhang, D.~S., Li, Q., Nan, J.~X., Liu, C.~Y., Bian, W., Ye, J.~P., and Acm (2018b).
\newblock Large-scale order dispatch in on-demand ride-hailing platforms: A learning and planning approach.
\newblock In {\em 24th ACM SIGKDD Conference on Knowledge Discovery and Data Mining (KDD)}, pages 905--913.

\bibitem[Yang et~al., 2020]{yang2020optimizing}
Yang, H., Qin, X., Ke, J., and Ye, J. (2020).
\newblock Optimizing matching time interval and matching radius in on-demand ride-sourcing markets.
\newblock {\em Transportation Research Part B: Methodological}, 131:84--105.

\bibitem[Yu and Shen, 2020]{yuIntegratedDecompositionApproximate2020}
Yu, X. and Shen, S. (2020).
\newblock An {{Integrated Decomposition}} and {{Approximate Dynamic Programming Approach}} for {{On-Demand Ride Pooling}}.
\newblock {\em IEEE Transactions on Intelligent Transportation Systems}, 21(9):3811--3820.

\end{thebibliography}

\newpage
\setcounter{table}{0}
\renewcommand{\thetable}{A.\arabic{table}}
\newpage

\section*{Appendix A: Notations}

\begin{longtable}{p{2cm}p{12cm}}
    \caption{Notations for the fleet dispatching and vehicle routing model}\label{tab: model} \\
    \hline
    Notations  & Explanation\\
    \hline
	\endfirsthead
	\hline
    Notations   & Explanation\\
    \hline
	\endhead
    \hline
    \multicolumn{2}{r@{}}{ }
    \endfoot
    \hline
    \endlastfoot
    \multicolumn{2}{l}{\emph{\textbf{Sets}}}\\
    $\mathcal U$ & the set of cities\\
    $\mathcal L$ & the set of intercity lines\\
    $\mathcal T$ & the set of dispatching horizons\\
    $\mathcal{K}^t_{u}$  & the set of available vehicles for assignment at horizon $t$ at city $u$\\
    $\hat {\mathcal{K}}^t_{u}$  & the set of vehicles that newly enter the system at horizon $t$ at city $u$\\
    $\tilde {\mathcal{K}}^t_u$  & the set of vehicles that exit the system at horizon $t$ at city $u$\\
    $\check {\mathcal{K}}^t_{u}$ & the set of arrived vehicles at horizon $t$ at city $u$\\
    $\bar  {\mathcal{K}}^{t}_{u}$ & the set of reserved vehicles at horizon $t$ at city $u$\\
    $\tilde {\mathcal{K}}^{t}_{u}$ & the set of vehicles that exit the system due to reaching limited work duration at horizon $t$ at city $u$\\
    $\mathcal N_{uv}^t$ & the set of vehicles dispatched from city $u$ to city $v$ at horizon $t$\\
    $\mathcal K_{uv}$ & the set of vehicles en route of line $(u,v)$\\
    $\mathcal O_{uv}$ & the set of orders in the matching pool and in service for line $(u,v)$\\
    $\mathcal O^1_{uv}$ & the set of orders for line $(u,v)$ that passengers have been picked\\
    $\mathcal O^2_{uv}$ & the set of orders for line $(u,v)$that have been matched but not picked\\
    $\mathcal O^3_{uv}$ & the set of orders in matching pool of line $(u,v)$\\
    $\mathcal V_k$ & the set of vehicle locations at the beginning of a matching interval\\
    $\mathcal V_f$ & the set of destination nodes of all orders\\
    $\mathcal V_s$ & the set of origin nodes of the unserved orders\\
    
    \multicolumn{2}{l}{\emph{\textbf{Parameters}}}\\
    $\bar W$ & maximum daily work time for a driver\\
    $R^{uv}$ & trip fare of line $(u,v)$\\
    $C$ & traveling cost per unit of distance\\
    $p_e$ & the penalty rate for losing one passenger\\
    $v_k$ & location of vehicle $k$ at the beginning of a matching interval\\
    $n_p$ & number of passengers of order $p$\\
    $s_p$ & origin node of order $p$\\
    $f_p$ & destination node of order $p$\\
    $[\check L_p, \check U_p]$ & time window of origin node of order $p$\\
    $[\hat L_p, \hat U_p]$ & time window of destination node of order $p$\\
    $k_p$ & matched vehicle of order $p\in O^1_{uv}\cup O^2_{uv}$\\
    $E$ & the location of the depot in the destination city\\
    $D_{ij}$ & the distance of arc $(i,j)$ in the graph for routing\\
    $W$ & the capacity of the vehicle\\
    $s$ & the speed of the vehicle\\
    
    \multicolumn{2}{l}{\emph{\textbf{Variables}}}\\
    $s_k^t$  & the number of passengers that vehicle $k$ picked in the trip begins at horizon $t$\\
    $\epsilon_{uv}^t$ & the number of lost passengers in line $(u,v)$ at horizon $t$\\
    $\delta^t_k$ & the total distance of the trip of vehicle $k$ begins at horizon $t$ \\
    $\epsilon^t_{uv}$ & the number of lost passengers in line $(u,v)$ at horizon $t$\\
    \multicolumn{2}{l}{\emph{\textbf{Decision variables}}}\\
    $x^{pk}_{ij}$   & binary variable indicating whether vehicle $k$ passes arc $(i,j)$ with passengers of order $p$\\
    $y^{k}_{ij}$ & binary variable indicating whether vehicle $k$ passes arc $(i,j)$\\
    $d_{pk}$ & binary variable indicating whether vehicle $k$ serves order $p$\\
    $u_{kj}$ & the time for vehicle $k$ to arrive at node $j$. The constraints can be formulated as follows 
\end{longtable}

\end{document}